\documentclass[useAMS,usenatbib]{mn2e}

\usepackage{graphicx}
\usepackage{longtable}
\pdfoutput=1

\title[Optical Observations of PSR\, J0205+6449 - the next optical pulsar?]{Optical Observations of PSR\, J0205+6449 - the next optical pulsar?}
\author[A. Shearer et al.]
{\parbox{\textwidth}{P. Moran$^{1}$, 
R. P. Mignani$^{2,3,4}$, 
S. Collins$^{1}$, 
A. de Luca $^{3,5}$, 
N.Rea$^{6}$, 
A. Shearer$^{1}$\thanks{E-mail:andy.shearer@nuigalway.ie (AS); rm2@mssl.ucl.ac.uk (RM)}}  \\ \\
$^{1}$ Centre for Astronomy, National University of Ireland, Newcastle Road, Galway, Ireland\\
$^{2}$ Mullard Space Science Laboratory, University College London, Holmbury St. Mary, Dorking, Surrey, RH5 6NT, UK\\
$^{3}$ INAF - Istituto di Astrofisica Spaziale e Fisica Cosmica Milano, via E. Bassini 15, 20133, Milano, Italy\\
$^{4}$ Kepler Institute of Astronomy, University of Zielona G\'ora, Lubuska 2, 65-265, Zielona G\'ora, Poland\\
$^{5}$ INFN - Istituto Nazionale di Fisica Nucleare, sezione di Pavia, via A. Bassi 6, 27100, Pavia, Italy \\
$^{6}$ Institut de Ci\'encies de l'Espai (IEEC-CSIC), Campus UAB, Fac. de Ci\'encies, Torre C5-parell, 2a planta, 08193 Barcelona, Spain 
}

\begin{document}

\def \psr{PSR\, J0205+6449}

\date{}

\pagerange{\pageref{firstpage}--\pageref{lastpage}} \pubyear{2002}

\maketitle

\label{firstpage}

\begin{abstract}
\psr\ is a young ($\sim$ 5400 years), Crab-like pulsar detected in radio and at X and $\gamma$-ray energies and has the third largest spin-down flux among known rotation-powered pulsars. It also powers a bright synchrotron nebula detected in the optical and X-rays. At a distance of $\sim 3.2$ kpc and with an extinction comparable to the Crab, \psr\ is an obvious target for optical observations. We observed \psr\ with several optical facilities, including 8m class ground-based telescopes, such as the Gemini and the Gran Telescopio Canarias. We detected a point source, at a significance of 5.5$\sigma$, of magnitude i'$\sim$25.5, at the centre of the optical synchrotron nebula, coincident with the very accurate {\em Chandra} and radio positions of the pulsar.
Thus, we discovered a candidate optical counterpart to \psr. The pulsar candidate counterpart is also detected in the g' ($\sim$27.4) band and weakly in the r' ($\sim$26.2)  band. Its optical spectrum is fit by a power law with photon index $\Gamma_{O} = 1.9 \pm 0.5$, proving that the optical emission if of non-thermal origin, is as expected for a young pulsar. The optical photon index is similar to the X-ray one ($\Gamma_{X}=1.77\pm 0.03$), although the optical fluxes are below the extrapolation of the X-ray power spectrum. This would indicate the presence of a double spectral break between the X-ray and optical energy range, at variance with what is observed for the Crab and Vela pulsars, but similar to the Large Magellanic Cloud pulsar PSR\, B0540$-$69. 

\end{abstract}


\section{Introduction}

PSR\, J0205+6449 in supernova remnant (SNR) 3C\,58  is a young energetic pulsar detected at X, $\gamma$-ray, and radio wavelengths \citep{2002ApJ...571L..41C, 2002ApJ...568..226M, 2009ApJ...699L.102A}. It has a spin period $P=$65 ms and  the third highest spin down energy flux, $\dot{E}/d^2 \approx 2.6~10^{36}$ erg s$^{-1}$ kpc$^{-2}$ (where $d$ is the pulsar distance), after the Crab and Vela pulsars.  3C\,58  was thought to be young, associated with supernova SN 1181, \citep{1978ApJ...220L...9V},  and consequently should share many of the characteristics of the Crab nebula, including the presence of a young pulsar. Yet its pulsar,  PSR\, J0205+6449, defied detection for over twenty years. It was only about ten years ago that it was discovered as an X-ray pulsar by \emph{Chandra} (Murray et al.\ 2002), while its identification as a radio pulsar came soon after (Camilo et al.\ 2002).  \psr\ was also detected in the hard X-rays (Kuiper et al.\ 2010) by the  {\em High Energy X-ray Timing Experiment} ({\em HEXTE}) aboard the  {\rm Rossi X-ray Timing Explorer} ({\em RXTE})  and was also one of the first to be identified as a $\gamma$-ray pulsar by {\em Fermi} (Abdo et al.\ 2009).  PSR\, J0205+6449  is clearly located within a pulsar wind nebula (PWN), detected both in the X-rays (Slane et al.\ 2002; 2004), optical \citep{2008A&A...486..273S}, and near-infrared (Slane et al.\  2008). 

From the X-ray determined hydrogen column density (Marelli et al.\ 2011; Marelli 2012), and using the \citet{1995A&A...293..889P} relation, we can estimate an  $A_V \sim 2.2$-2.5, similar to the Crab pulsar.  The distance to \psr, however, is possibly larger than the Crab.  HI  radio observations gives a distance to 3C\,58 of  3.2 kpc  \citep{1993A&A...274..427R}, lower but still consistent with the value of  $4.5^{+1.58}_{-1.21}$ kpc obtained from the pulsar Dispersion Measure of $140.7\pm0.3$ cm$^{-3}$ pc \citep{2002ApJ...571L..41C} and the electron distribution model of \citet{2002astro.ph..7156C}. 

If the Crab and 3C\,58 SNRs were of the same age, we might also expect similarities between their pulsars.   However, PSR\, J0205+6449 was found to be considerably weaker than the Crab pulsar; its X-ray emission is 1000 times lower than the Crab and its radio emission is 120 times lower. In $\gamma$-rays, PSR\, J0205+6449's  luminosity is about 10\% of the Crab pulsar's but its efficiency, for an assumed 3.2 kpc distance, is about 0.2--0.3\% compared to the Crab's 0.1\% \citep{A2010} consistent with PSR\, J0205+6449 being an older pulsar.   Some of this discrepancy can be attributed to the mounting evidence that the characteristic age of the pulsar  (given by $P \over{2\dot{P}} $ $\approx$ 5400 years) is near to its true age \citep{2005ApJ...619..839C, 2006ApJ...645.1180B, 2007ApJ...654..267G},  with the age estimates for the  3C\,58  SNR  ranging between 3000 and 5100 years, thus breaking the association between supernova SN1181 and \psr/3C\,58. Recently, however, a re-analysis of the existing HI radio observations (Kothes\, 2010) suggests a distance as small as 2 kpc for 3C\, 58 and an age as low as $\sim$ 1000 years, reopening the debate on its association with SN1181.

The  multi-wavelength properties, age, and energetics of \psr\ combine to make  it a likely candidate for optical emission studies. If we assume that the pulsar's optical luminosity scales with the light cylinder magnetic field $B_{lc}^{1.6}$ \citep{2001ApJ...547..967S}, then we estimate that it should have a visual magnitude in the range 23--25, depending on interstellar absorption and effects of beaming geometry.  The first deep observations of 3C\,58 \citep{2008A&A...486..273S,2008MNRAS.390..235S} showed evidence of an optical nebulosity at the same location as the X-ray counterpart to PSR\, J0205+6449 but could not resolve the pulsar optical counterpart. The motivation behind this work was to identify a candidate optical counterpart to PSR\, J0205+6449 using observations taken under the 2009 International Time Programme at the La Palma Observatory (PI: A. Shearer), combined with observations recently obtained with the 10.4m Gran Telescopio Canarias (GTC) as well as archival Gemini and  {\em Hubble Space Telescope} ({\em HST}) images. Table 1 reports the summary of the \psr\  observations with the different facilities.

\section[]{Observations and data reduction}

\subsection{Isaac Newton Telescope}

We first observed the \psr\ field  with the 2.5m UK Isaac Newton Telescope (INT) at the  La Palma Observatory (Roque de Los Muchachos, Canary Islands, Spain) on 2009 August 3. We used the Wide Field Camera (WFC), a mosaic of 4 thinned EEV CCDs, with an unbinned  pixel size of  0\farcs33 and with a field--of--view of $34\farcm2\times34\farcm2$, including  the $\sim$ 1\arcmin gap between the chips. We only observed through the Harris R-band filter ($\lambda=6380$\AA; $\Delta \lambda=1520$\AA) for a total integration time of 7320 seconds, with an average airmass of 1.35 and a seeing of 1\farcs3. Observations were performed in grey time, with the Moon at an angular distance of $\sim 110^{\circ}$ but mostly below the horizon at the moment of the observations. The integration was split in shorter dithered exposures to remove cosmic ray hits and compensate for the gaps between the CCD chips.  

\subsection{Telescopio Nazionale Galileo}

On 2009 August 20 and 21, we re-observed the \psr\  field with the 3.5m italian Telescopio Nazionale Galileo (TNG), also at the La Palma Observatory. We used the  DOLORES (Device Optimized for the LOw RESolution) camera,  a single chip E2V CCD with a field--of-view of $8\farcm6 \times 8\farcm6$ and a pixel size of 0\farcs252. We observed through the standard Johnson V ($\lambda=5270$\AA; $\Delta \lambda=980$\AA) and R ($\lambda=6440$\AA; $\Delta \lambda=1480$\AA) filters for total integration times of 6060 and 4830 seconds, respectively.  In both cases, the average airmass was around 1.5 and the seeing $\sim 1\farcs3$. Observations were performed in dark time. For both the INT and TNG images, we applied standard data reduction (bias subtraction and flat-fielding) using the tools in the {\sc IRAF} packages {\sc mscred} and  {\sc ccdred}. The dithered exposures were then aligned, stacked, and filtered for cosmic-rays so as to produce mosaic images.

\subsection{Gran Telescopio Canarias}

We obtained additional observations of the \psr\ field with the GTC  at the  La Palma Observatory on 2011 September 1 and 2011 November 20 as part of the Spanish Time programme (PI: N. Rea). We observed \psr\ with the Optical System for Imaging and low Resolution Integrated Spectroscopy (OSIRIS). The instrument is equipped with a two-chip E2V CCD detector with a nominal field--of--view of $7\farcm8\times8\farcm5$, which is actually decreased to $7\arcmin \times 7\arcmin$ due to the vignetting of one of the two chips. The unbinned pixel size of the CCD is 0\farcs125. We took a sequence of dithered exposures in the SDSS r' ($\lambda=6410$\AA; $\Delta \lambda=1760$\AA) and i' ($\lambda=7705$\AA; $\Delta \lambda=1510$\AA) bands on the first and second night, respectively,  with exposure time of 140 seconds  to minimise the saturation of bright stars in the field and correct for the fringing. The pulsar was positioned at the nominal aim point in chip 2. Observations were performed with an average airmass of 1.25 for both the r' and i' bands. Seeing conditions were $\sim 0\farcs9$  and $0\farcs8$--1\farcs0 for the first and second night, respectively.  In both nights, observations were performed in dark time and under clear conditions.  For the i'-band observations, no valid sky-flats were taken for the night of November 20, therefore we used closest--in--time sky-flats taken on November 17.  As done for the INT and TNG data, we reduced the data using standard tools in the {\sc IRAF} package {\sc ccdred}. We then stacked and averaged the single dithered exposures using the task {\tt drizzle} that also performs the cosmic-ray filtering.

\begin{table}
\caption{Observation log for the PSR\, J0205+6449 observations}
\begin{center}
\begin{tabular}{l||c|c|r|c|}
\hline
Telescope & Date & No. of     &Integration & Filter  \\
                  &           & Frames & Time (s) & \\
\hline
INT       	&2009 Aug 03 	& 9 	& 2520 		& Harris R \\
             	&2009 Aug 03 	& 4 	& 4800 		& Harris R \\ \hline
TNG     	&2009 Aug 20 	& 7 	& 3060 		& Johnson V \\
             	&2009 Aug 20 	& 5 	& 3000 		& Johnson V \\
             	&2009 Aug 21 	& 6 	& 4830 		& Johnson R \\ \hline
GTC    		&2011 Sep 01 	& 5  	& 700  		& r' \\
            	&2011 Nov 20 	& 17 	& 2380		& i' \\      \hline            
Gemini 		&2007 Aug 10 	& 4 	& 3600 		& r\_G0303 \\
            	&2007 Aug 10 	& 4 	& 3600 		& g\_G0301 \\     
            	&2007 Aug 10 	& 2 	& 1800 		& i\_G0302 \\     
            	&2007 Sep 06 	& 1 	& 30 		& i\_G0302 \\     
            	&2007 Sep 11 	& 3 	& 2700 		& i\_G0302 \\     
            	&2007 Sep 14 	& 1 	& 900 		& i\_G0302 \\ 
            	&2007 Sep 14 	& 3 	& 2700 		& r\_G0303 \\ 
            	&2007 Oct  03  	& 1 	& 900 		& r\_G0303 \\ 
            	&2007 Oct 03  	& 2 	& 1800 		& g\_G0301 \\ 
           	&2007 Oct 09  	& 2 	& 120 		& g\_G0301 \\ 
           	&2007 Oct 09  	& 2 	& 120 		& r\_G0303 \\ 
           	&2007 Oct 09  	& 2 	& 120 		& i\_G0302 \\ 
           	&2007 Oct 09  	& 1 	& 900 		& r\_G0303 \\ 
           	&2007 Oct 09  	& 2 	& 1800 		& g\_G0301 \\ 
            	&2007 Oct 10 	& 1 	& 900 		& g\_G0301 \\ 
            	&2007 Oct 11 	& 3 	& 2700 		& g\_G0301 \\ \hline
HST	     	&2009 Nov 27 	& 3 	& 1800 		& 625W \\
             	&2009 Nov 27 	& 3  	& 1950 		& 775W \\ \hline
\end{tabular}
\end{center}
\label{obs}
\end{table}

\subsection{Gemini}

Using the {\em Gemini} science data archive\footnote{{\tt http://cadcwww.dao.nrc.ca/gsa/}}  we identified 34 frames of the PSR\, J0205+6449 field taken between 2007 August 10 and 2007 October 11 with the Gemini-North telescope in Mauna Kea (Hawaii). Observations were performed using the Gemini Multi-Object Spectrograph (GMOS). At the time of the observations the instrument was still mounting the original  three-chip EEV CCD detector that has a field--of--view of $5\farcm5 \times 5\farcm5$, with gaps  of $2\farcs8$ between each chip, and a pixel scale of 0\farcs1454. Observations were performed through the g\_G0301 ($\lambda=4750$\AA; $\Delta \lambda=1540$\AA), r\_G0303 ($\lambda=6300$\AA; $\Delta \lambda=1360$\AA), and i\_G0302 ($\lambda=7800$\AA; $\Delta \lambda=1440$\AA) filters, very similar to the g', r', and i' used by the Sloan Digitised Sky Survey (SDSS; Fukugita et al.\ 1996).   In total, the data corresponds to 10920 seconds integration time in the g' band, 8220 seconds in r', and 5521 seconds in i'. The average airmass during the observations was between 1.47 and 1.62 and the seeing between  0\farcs57 and 0\farcs73. Observations were all performed in dark time.

We reduced the GMOS images using the dedicated  {\sc gmos}  image reduction package available in {\sc IRAF}. After  downloading the closest--in--time bias and sky flat field frames from the {\em Gemini}  science archive, we used the  tasks {\tt gbias} and {\tt giflat} to process and combine the bias and  flat frames, respectively.  We  then reduced the single science frames using the task {\tt gireduce} for bias subtraction, overscan correction, image trimming and flat field normalisation.  From  the  reduced science images, we produced a mosaic of the three GMOS CCDs using the task {\tt  gmosaic} and we average-stacked the reduced image mosaics with the task  {\tt imcoadd} to filter out cosmic ray hits.

\subsection{Hubble Space Telescope}

Images of the \psr\ field are also available in the {\em HST} archive\footnote{{\tt http://archive.stsci.edu/}} (Program 11723). The observations were performed  on 2009 November 27 with the Wide Field Camera 3 (WFC3) and the UVIS detector, which has a field--of--view of $162\arcsec \times 162\arcsec$ and a pixels size of 0\farcs04. Images were obtained through the broad-band 625W ($\lambda=6250$\AA; $\Delta \lambda=1550$\AA) and 775W  ($\lambda=7760$\AA; $\Delta \lambda=1470$\AA) filters, similar to the SDSS filters r and i,  for a total integration time of 1800 and 1950 seconds, respectively. We retrieved the data from the {\em HST} archive, after on--the--fly recalibration by the WFC3 pipeline ({\sc CALWF3} version 2.3) that applies bias subtraction and flat field correction and produces distortion-corrected and co-added images.

\begin{figure}
\begin{center}
\includegraphics[height=60mm]{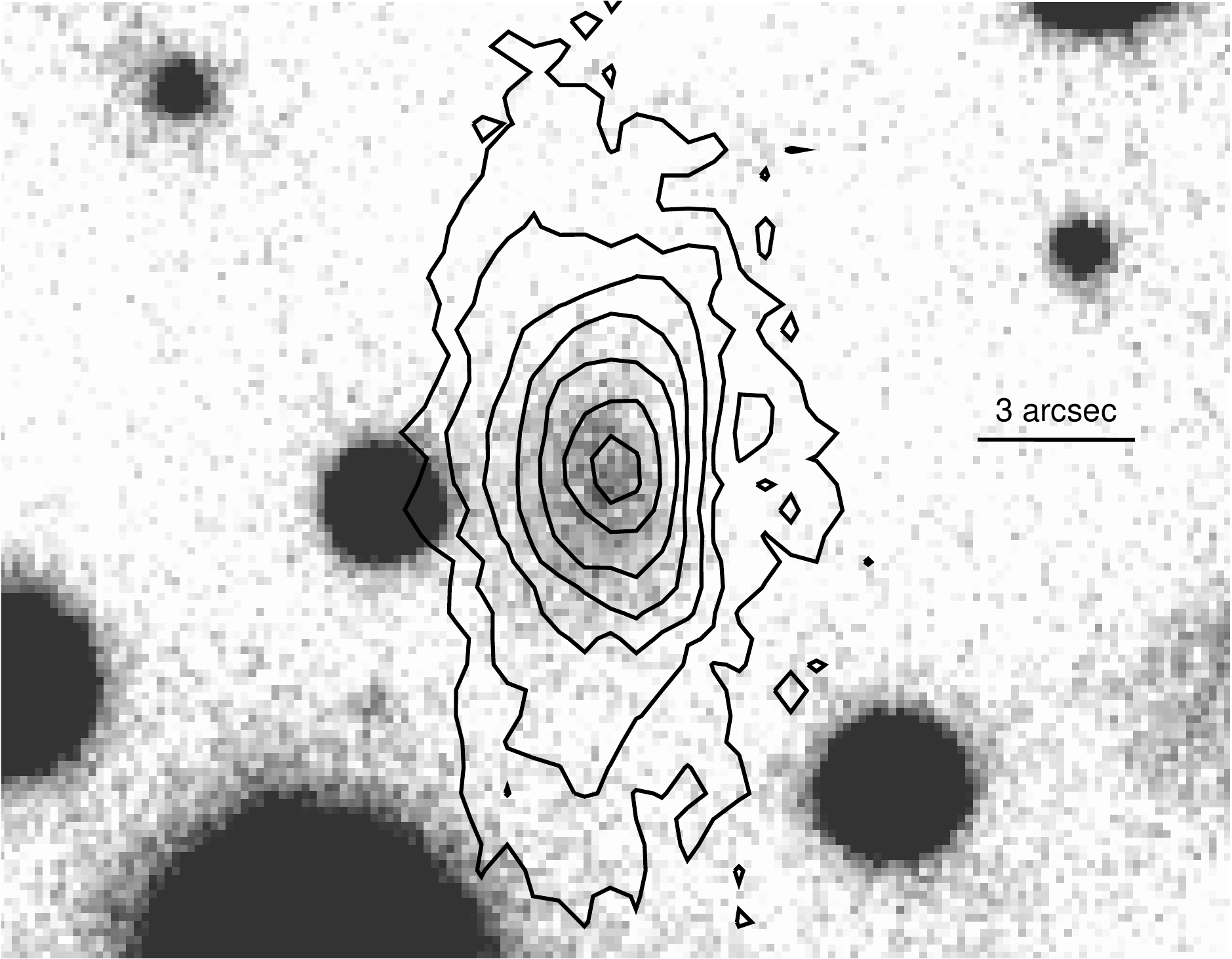}
\includegraphics[height=60mm]{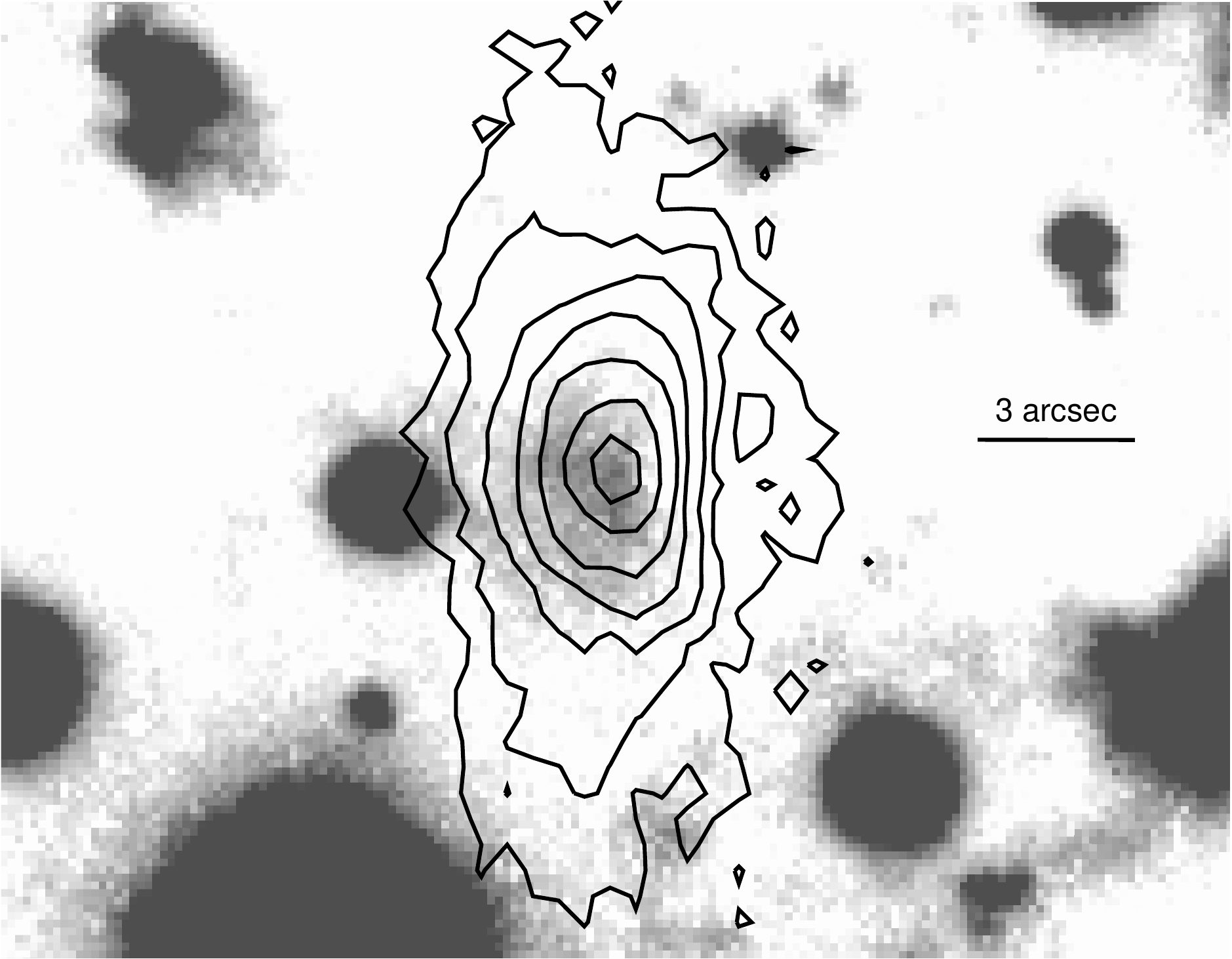}
\includegraphics[height=60mm]{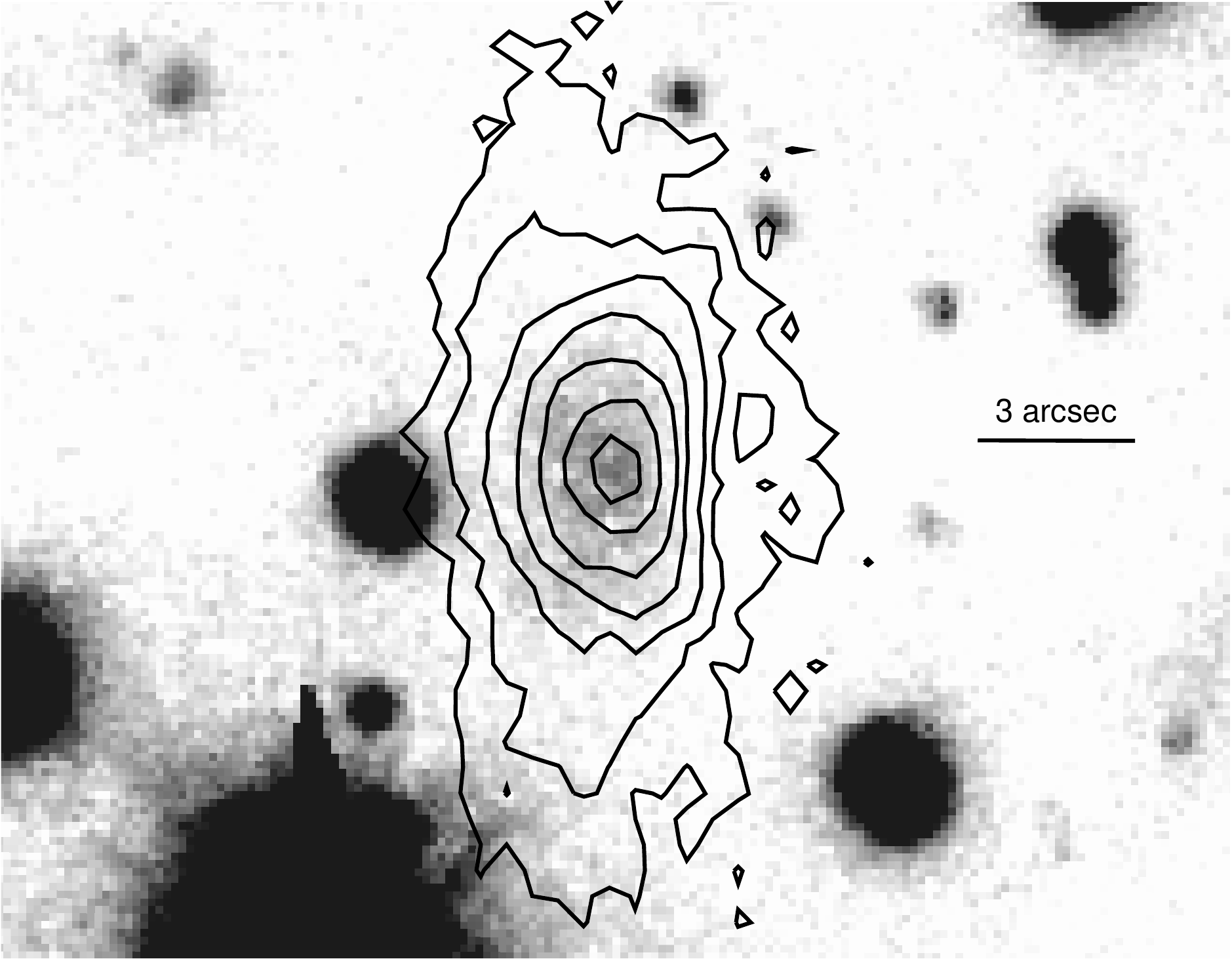}
\caption{High contrast Gemini images of the \psr\ field: g' (top), r'(middle) and i'(bottom). The {\em Chandra} X-ray contours are overlaid in black.}
\label{Gem_gri}
\end{center}
\end{figure}

\begin{figure}
\begin{center}
\includegraphics[height=60mm]{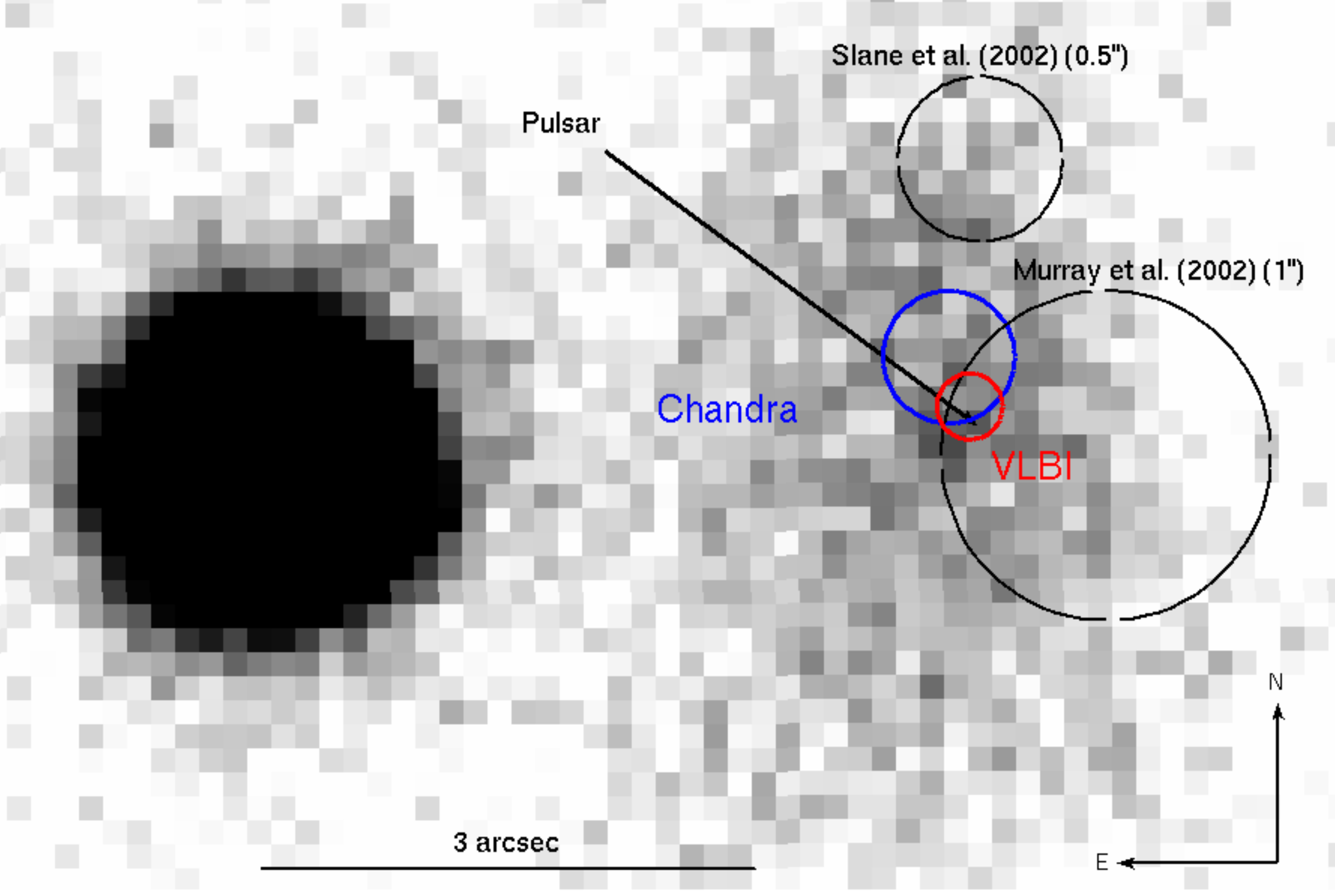}
\includegraphics[height=60mm]{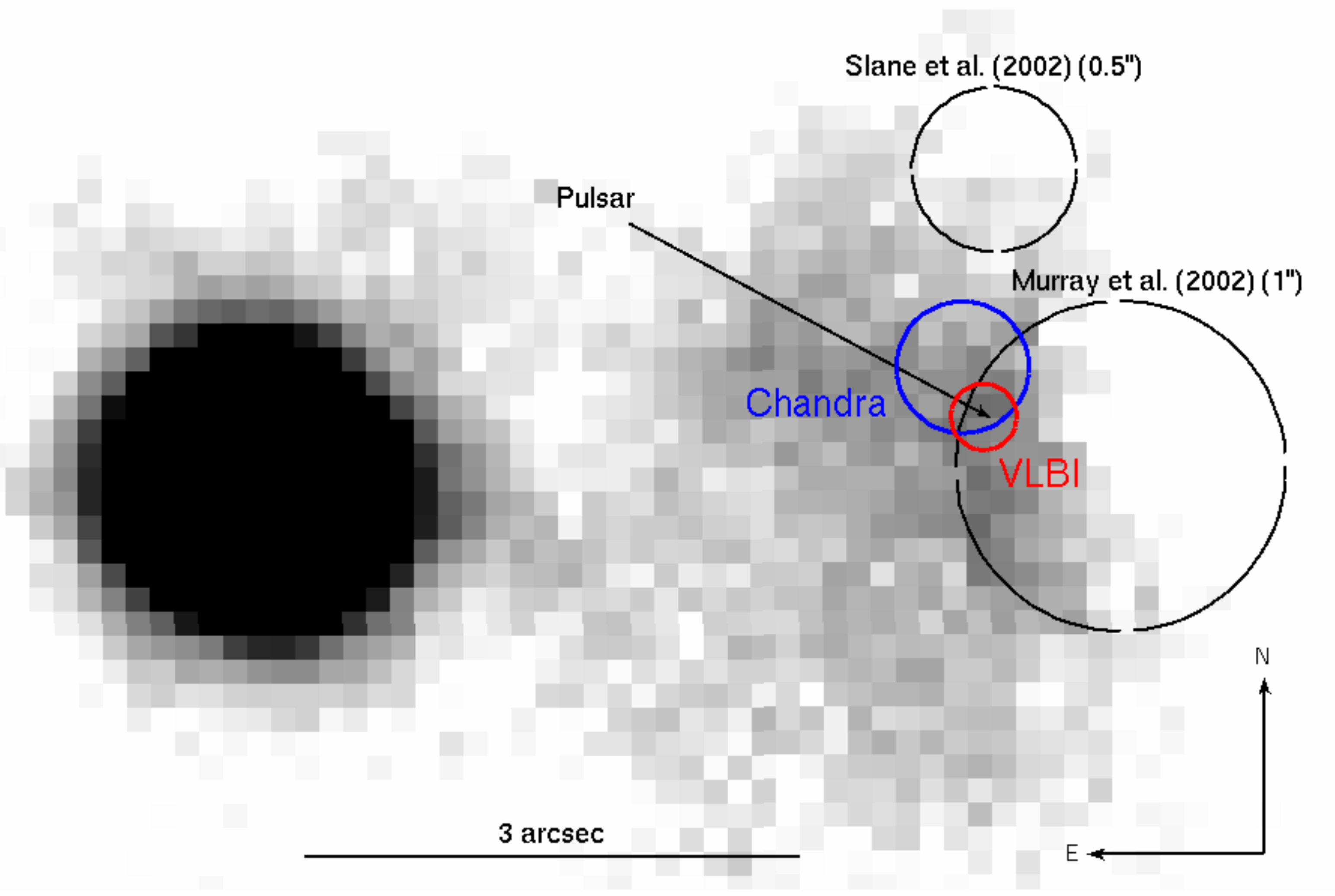}
\includegraphics[height=60mm]{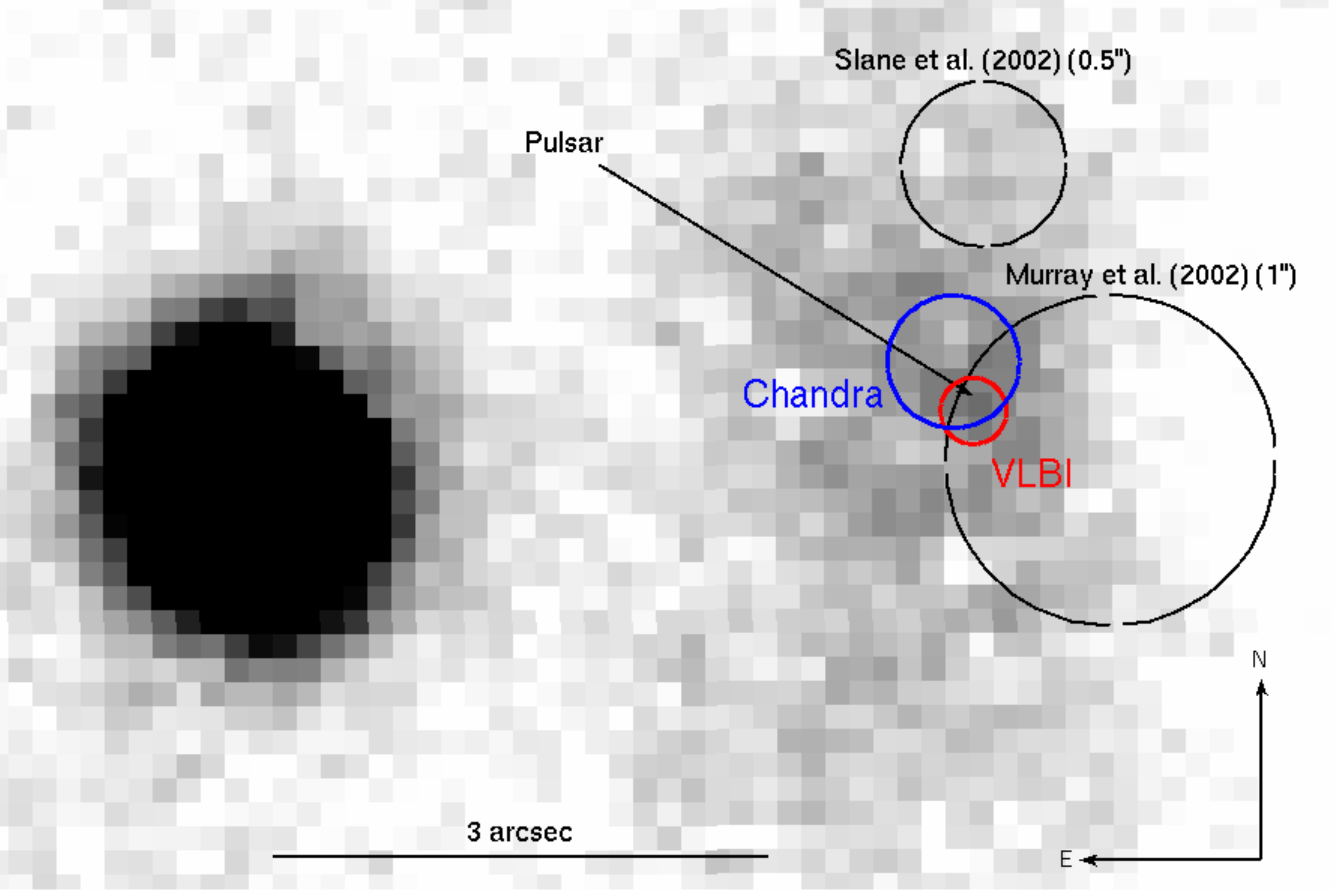}
\caption{High contrast Gemini images of the \psr\ field: g' (top), r' (middle), i' (bottom). The Slane et al.\ (2002) and Murray et al.\ (2002) {\em Chandra} error circles of the pulsar are overlaid (in black), together with updated {\em Chandra} error circle (this work; blue) and the Green Bank/VLBI radio error circle (Bietenholz et al.\ 2013; red). The pulsar candidate counterpart is marked by the arrow. We note that the systematic uncertainty on the radio position of Bietenholz et al.\ (2013) is only 10 mas (Bietenholz priv. comm.).  Thus, the radius of the radio error circle is dominated by the uncertainty of our optical astrometry.}
\label{figure2}
\end{center}
\end{figure}

\begin{figure}
\begin{center}
\includegraphics[width=75mm]{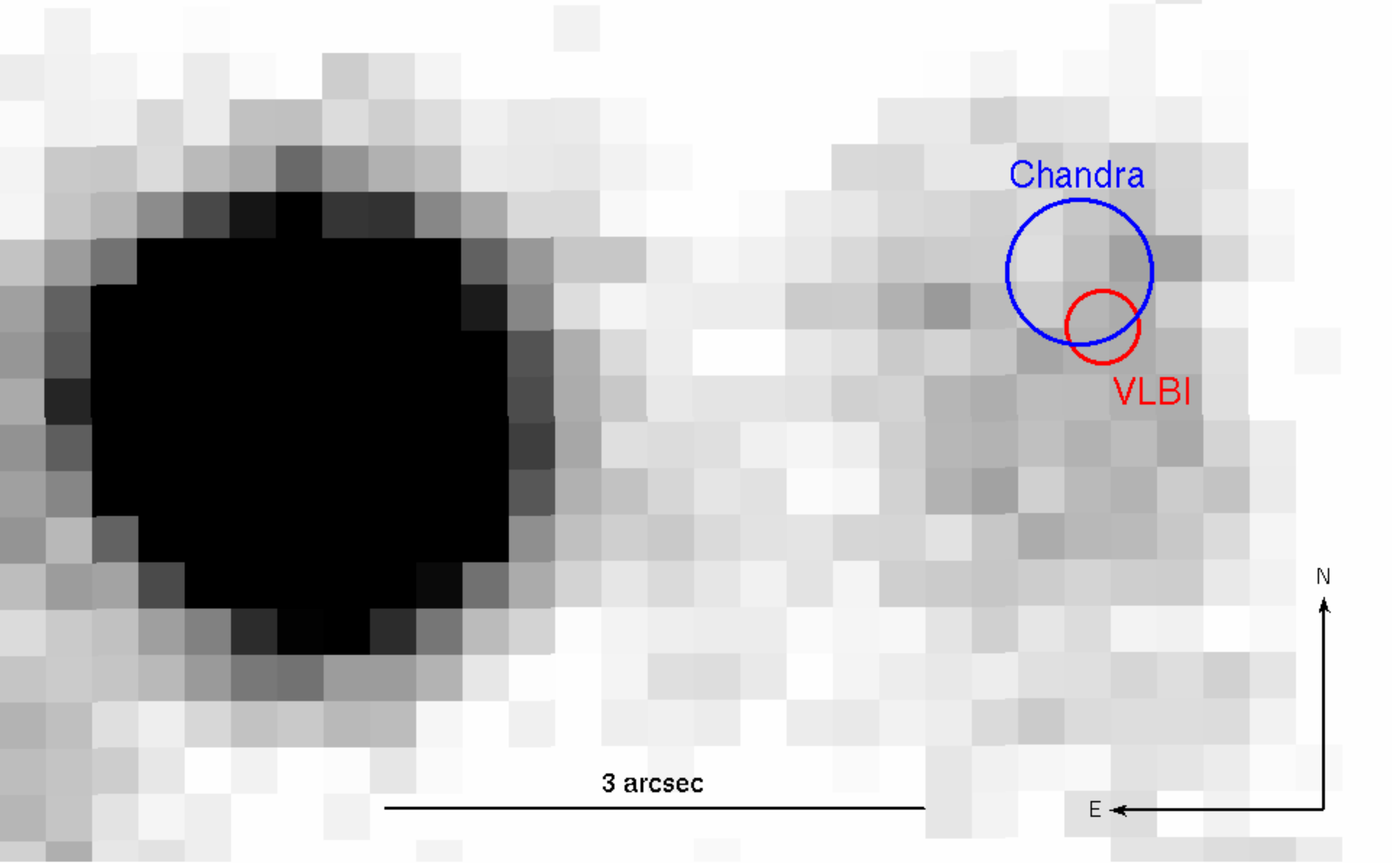}
\includegraphics[width=75mm]{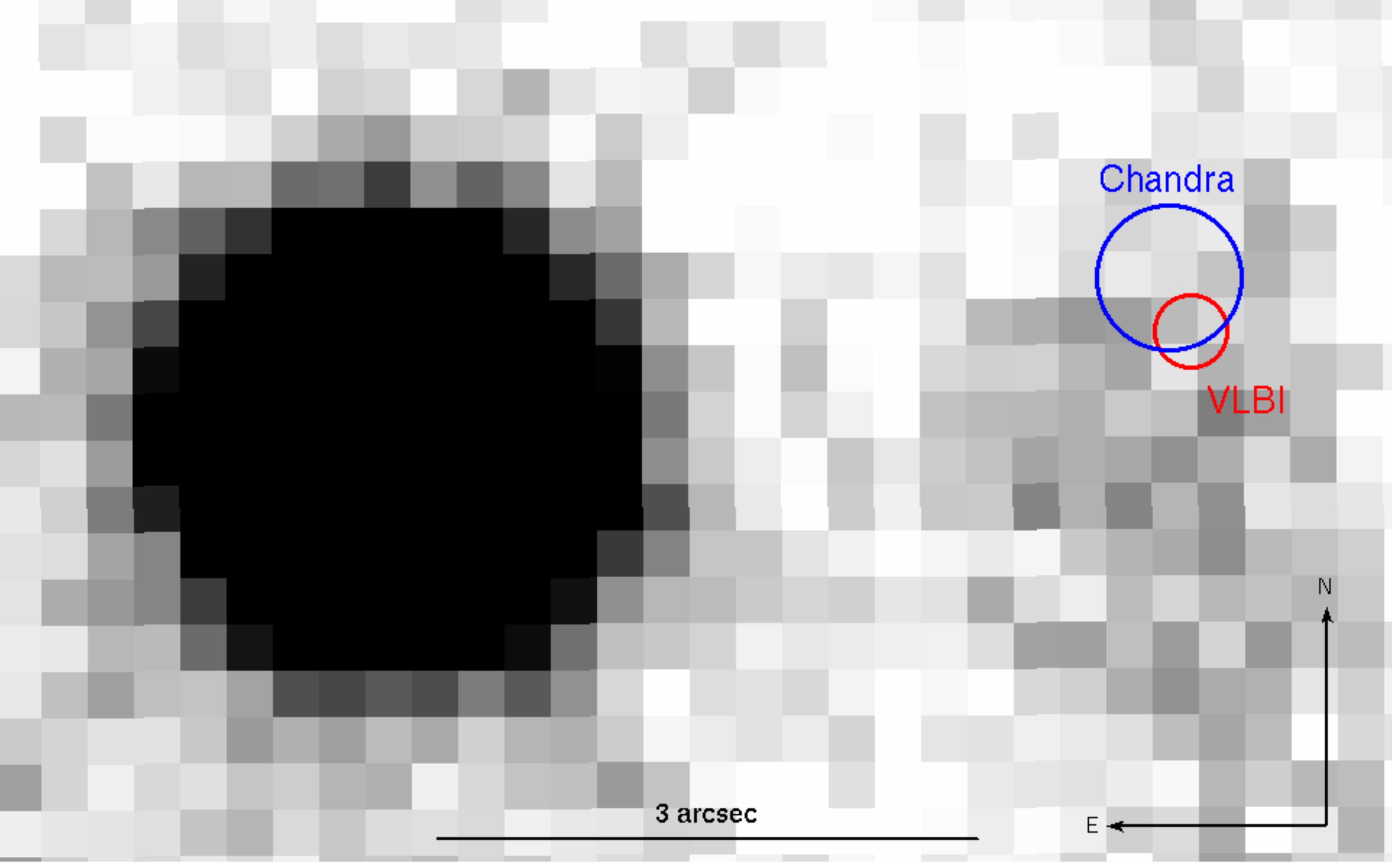}
\caption{GTC images: r' (top), and i' (bottom) images. The {\em Chandra} and  VLBI positions (Bietenholz et al.\  2013) are overlaid in blue and red, respectively.}
\label{figure3}
\end{center}
\end{figure}

\begin{figure}
\begin{center}
\includegraphics[width=75mm]{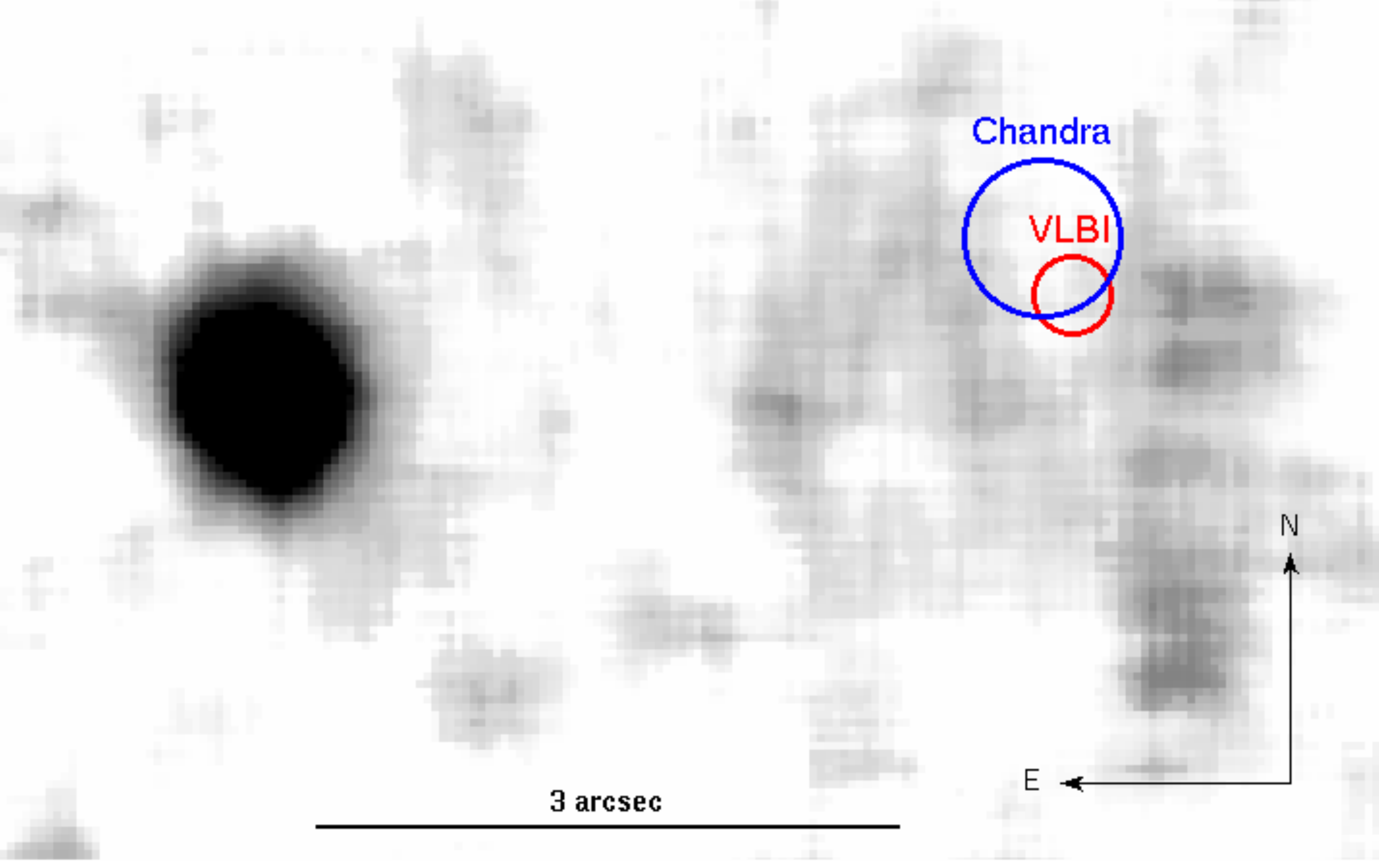}
\includegraphics[width=75mm]{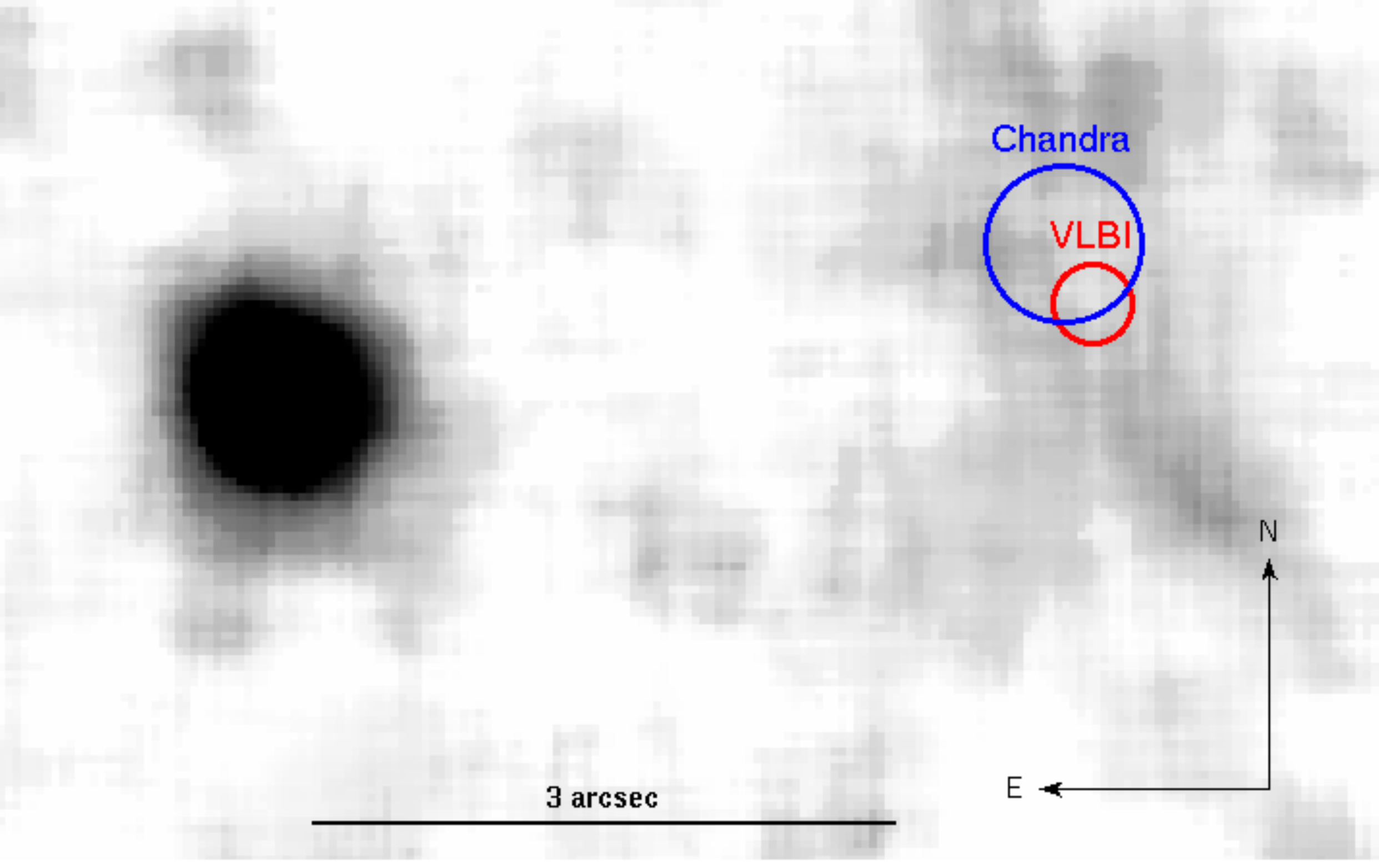}
\caption{{\em HST} images: F625W (top), and F775W (bottom). The {\em Chandra} and  VLBI  (Bietenholz et al.\  2013) positions are overlaid in blue and red, respectively. The PWN is visible but only because these images have been convolved with a median filter of 17 pixels.}
\label{figure4}
\end{center}
\end{figure}

\begin{figure}
\begin{center}
\includegraphics[width=80mm]{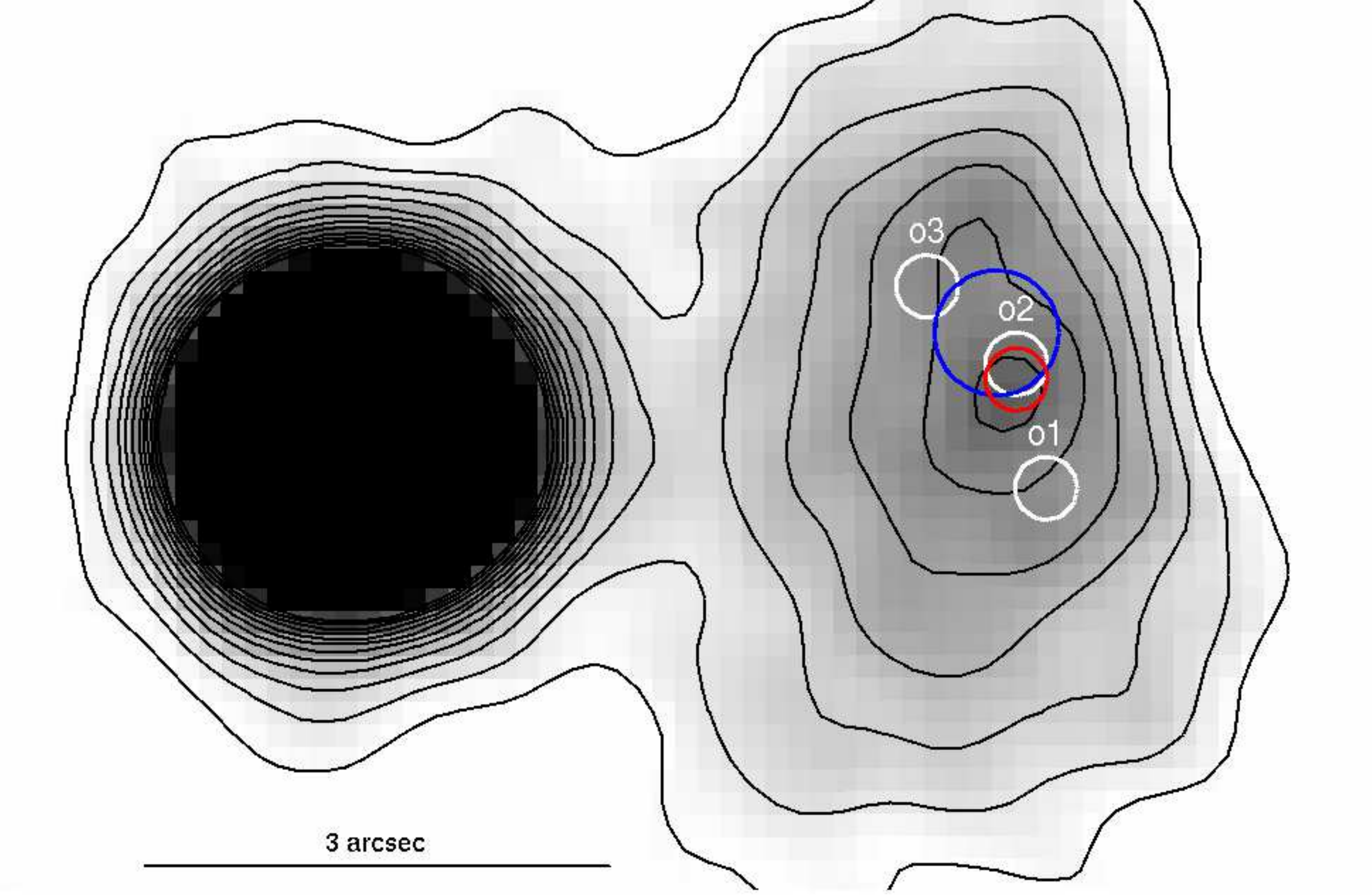}
\includegraphics[width=80mm]{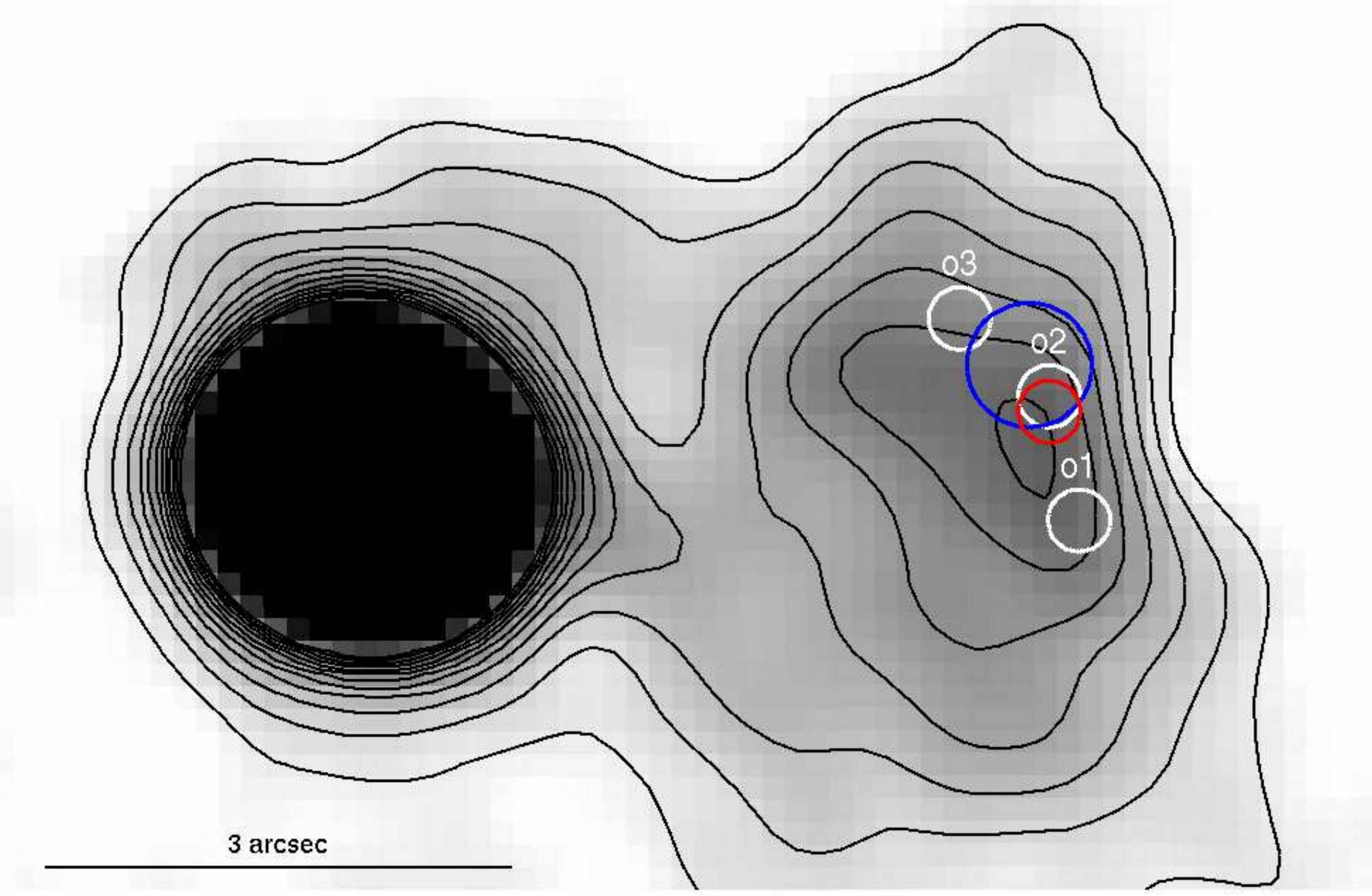}
\includegraphics[width=80mm]{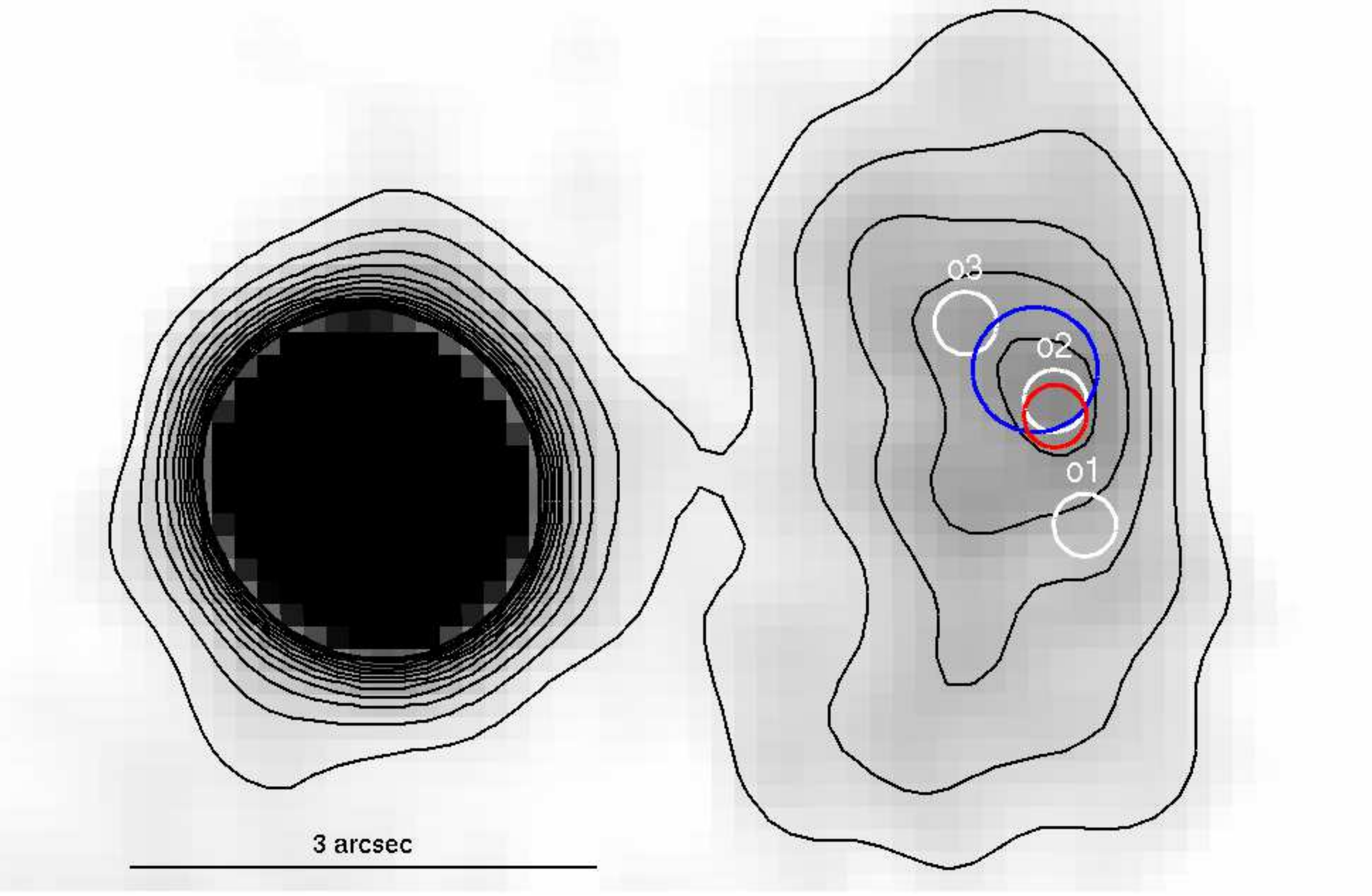}
\caption{High contrast Gemini images of the \psr\ field: g' (top), r'(middle) and i'(bottom). The images were smoothed with a Gaussian filter and optical contours are overlaid. The {\em Chandra} and VLBI (Bietenholz et al.\ 2013) positions of the pulsar are overlaid in blue and red, respectively. The positions of the filamentary structures (knots o1, o2, and o3) observed by Shearer \& Nuestroev (20008) are also marked.}
\label{figure5}
\end{center}
\end{figure}

\section{Astrometry}

\subsection{X-ray astrometry}

The position of PSR\, J0205+6449 has been given by two sources; the original discovery paper \citep{2002ApJ...568..226M} and subsequent analysis of the thermal properties of the pulsar \cite{2002ApJ...571L..45S}. In the former, there is a detailed analysis of the morphology of the central point source and PWN, yielding a position  $\alpha =02^{\rm h}  05^{\rm m} 37\fs8$ and  $\delta  = +64^\circ 49\arcmin 41$ with a positional error estimation of 1\arcsec. In the latter, based on relative astrometry with respect to the stellar counterparts of four field point sources, they derived offsets of $\Delta \alpha =-0\farcs17\pm0\farcs12$ and $\Delta\delta =0\farcs64\pm0\farcs08$ in the {\em Chandra} absolute astrometry. Then, they obtained $\alpha =02^{\rm h}  05^{\rm m} 37\fs92$ and  $\delta  = +64^\circ 49\arcmin 42\farcs8$ but with no estimation of fitting errors around the  pulsar.   \psr\ was also detected in the radio with the 100 m Green Bank Telescope (Camilo et al.\ 2002) but no position was independently measured. While the \cite{2002ApJ...568..226M} position was assumed as a reference both by \cite{2008A&A...486..273S} and \cite{2008MNRAS.390..235S} in their optical studies of the PWN,  the Slane et al.\ (2002) position was used, e.g. both by Livingstone et al.\  (2009) and \cite{2009ApJ...699L.102A} as a reference for the X and $\gamma$-ray timing analysis of the pulsar. No new position of \psr\  has been reported by the {\em Fermi} Pulsar Timing Consortium (Smith et al.\  2008).  Thus, since the actual \psr\ position is quite uncertain, and never independently re-assessed so far, we firstly re-computed it from the available {\em Chandra} observations.

We recomputed the X-ray position of PSR\, J0205+6449 using the two deepest \emph{Chandra} observations of the field (datasets 4382 and 3832, integration times 171 ks  and 139 ks, respectively), performed on April 23 and 26, 2003 with the ACIS-S detector. No more recent and comparably deep observations of \psr\ have been performed with {\em Chandra}. We retrieved the data from the public \emph{Chandra} Science Archive\footnote{{\tt http://cxc.harvard.edu/cda/}} and analysed them with the \emph{Chandra} Interactive Analysis of Observations (CIAO) software v.4.1.1. For each dataset, we extracted an image in the 0.5--6 keV energy range using the original ACIS pixel size ($0\farcs492$). We ran a source detection using the {\tt wavdetect} task with wavelet scales  ranging from 1 to 16 pixels, spaced by a factor of  $\sqrt{2}$ (consistent results were obtained  using the {\tt celldetect} task). To check the  accuracy of the \emph{Chandra}/ACIS absolute astrometry,  we selected X-ray sources detected at $>4 \sigma$  in both observations within 4\arcmin\ from the aim point, since the accuracy of the  \emph{Chandra} astrometry rapidly degrades at large off-axis angles. Then, we cross-correlated their positions with astrometric catalogues. We found 4 matches with the Two Micron All-Sky Survey (2MASS; Skrutskie et al.\ 2006) with offsets of 0\farcs1--0\farcs3, in agreement with the expected accuracy of the ACIS-S absolute astrometry\footnote{{\tt http://cxc.harvard.edu/cal/ASPECT/celmon/}}.  Three of such sources were also identified in the United States Naval Observatory (USNO) B1.0 catalogue (Monet et al.\ 2003),  with similar X-ray--to--optical offsets. Based on such coincidences,  we could assess that no systematic offsets affect the two astrometric solutions, although it was not possible to improve them. 

We computed the best position for PSR\, J0205+6449 by averaging the coordinates measured in the two observations. The resulting position (J2000) is $\alpha =02^{\rm h}  05^{\rm m} 37\fs95$ and  $\delta  = +64^\circ 49\arcmin 41\farcs6$ (epoch 2003.31), with a nominal radial uncertainty of $0\farcs375$  at the $90\%$ confidence level.  Thus, our re-computed position falls $\sim 0\farcs8$ South of the Slane et al.\ (2002) one, and is more consistent with the \cite{2002ApJ...568..226M} position, which is much closer to the centre of the optical PWN. 

While we were close to submit our manuscript, a {\em Chandra} position for \psr\ was published by Bietenholz et al.\ (2013), based on the very same data sets and perfectly consistent with ours.  Bietenholz et al.\ (2013) also reported on a new VLBI radio position of the pulsar, which is consistent with the {\em Chandra} one, and on the first measurement of the pulsar proper motion ($\mu_{\alpha}=1.4 \pm 0.16$ mas yr$^{-1}$; $\mu_{\delta}= 0.540\pm 0.575$ mas yr$^{1}$. Due to the small angular displacement between the epochs of the {\em Chandra} and VLBI positions and those of our optical observations, in the following we neglect the effect of the proper motion of the assumed pulsar position. We note that the systematic uncertainty on the radio position of Bietenholz et al.\ (2013) is only 10 mas (Bietenholz priv. comm.). Thus, the radius of the radio error circle is dominated by the uncertainty of the astrometric calibration of the optical images ($\sim0\farcs2$).

\subsection{Optical astrometry}

We computed the astrometry calibration of the optical images using the {\em wcstools}\footnote{{\tt http://tdc-www.harvard.edu/wcstools/}} suite of programs that automatically match the sky coordinates of stars in the selected reference catalogue with their pixel coordinates computed by {\em Sextractor} (Bertin \& Arnouts 1996).  In order to avoid systematics with the {\em Chandra} astrometry (see Sectn. 3.1), we used 2MASS as a reference catalogue.   After iterating the matching process and applying a sigma-clipping selection to filter out obvious mismatches, high-proper motion stars, and false detections, a pixel--to--sky coordinate transformation was computed using a polynomial function and we obtained,  for the ground-based images mean residuals of $\sim 0\farcs2$ in the radial direction, using 50 bright, but non-saturated, 2MASS stars. To this value we added in quadrature the uncertainty $\sigma_{tr}$ = 0\farcs08 of the image registration  on the  2MASS reference frame. This is given by $\sigma_{tr}$=$\sqrt{n/N_{S}}\sigma_{\rm S}$ (e.g., Lattanzi et al.\ 1997), where $N_{S}$ is the number of stars used to compute the astrometric solution, $n$=5 is the number of free parameters in the sky--to--image transformation model, $\sigma_{\rm S} \sim 0\farcs2$ is the mean absolute position error of  2MASS  for stars in the magnitude range  $15.5 \le K \le 13$ (Skrutskie et al.\ 2006). After accounting for the 0\farcs015  uncertainty on the link of 2MASS to the International Celestial Reference Frame  (Skrutskie et al.\ 2006), we ended up with an overall accuracy of $\sim$0\farcs22 on the absolute optical astrometry of the ground-based images. For the {\em HST} ones, thanks to their much better spatial resolution, we could measure the stars' relative position with a much better accuracy and obtained mean residuals of $\sim 0\farcs05$ on the pixel--to--sky coordinate transformation. This corresponds to an overall accuracy of $\sim 0\farcs1$ on the absolute astrometry of the {\em HST} images.

\section{Data analyses and results}

\subsection{Image analyses}

The deepest Gemini, GTC, and {\em HST} stacked images of the \psr\ field are shown in the Figures 2, 3, and 4 respectively, with the computed {\em Chandra} and VLBI \citep{2013...1302.5625} pulsar positions  marked. The optical PWN is visible in all the deepest ground-based images, including the INT and TNG ones, albeit with a different signal--to--noise due to the difference in integration time, observing conditions, and telescope/instrument sensitivity. The PWN is also visible in the highest resolution {\em HST} WFC3 images (Figure 4), but only when they are convolved with a median filter of 17 pixels. As a result, this smears the WFC3 resolution to the level of the ground-based optical images and prevents the study of the finer details in the PWN structure. The morphology of the optical PWN is better resolved in the deeper Gemini images (Figure 1), where its brightness distribution shows a clear central maximum. The nebula features an elongated morphology stretching North to South with a structure similar to that of the X-ray PWN. However, it extends on a much smaller angular scale ($\sim 6\arcsec$) and overlaps only the brightest part of the X-ray PWN.  Furthermore, there is no evidence of an optical counterpart of the curved, jet-like X-ray structure protruding west of the pulsar and detected in the {\em Chandra} image. 

According to our astrometry, the pulsar position is now closer to the centre of the optical PWN. In particular, the pulsar position is consistent with that of the emission maximum of the optical PWN, more clearly detected in the Gemini g'-band image (Figure 2, top), where  the presence of a faint, point-like, object becomes apparent.  This is also shown in Figure 1, where the position of this object is consistent with the centroid of the X-ray contours of the PWN.  This object is also detected in the Gemini i'-band image (Figure 2, bottom) and in the r'-band one (Figure 2, middle),  albeit at lower significance. Its independent detection in different  images, where it is in different positions of the detector, confirms that the object is not spurious and not produced either by CCD  blemishes or artefacts in the data reduction.   The object is not detected in the other ground-based images (Figure 3), owing to their lower sensitivity and the difficulty of resolving the structure of the PWN, and is not detected in the {\em HST} images, even after the images had been median filtered (Figure 4).

\begin{figure}
\begin{center}
\includegraphics[width=70mm]{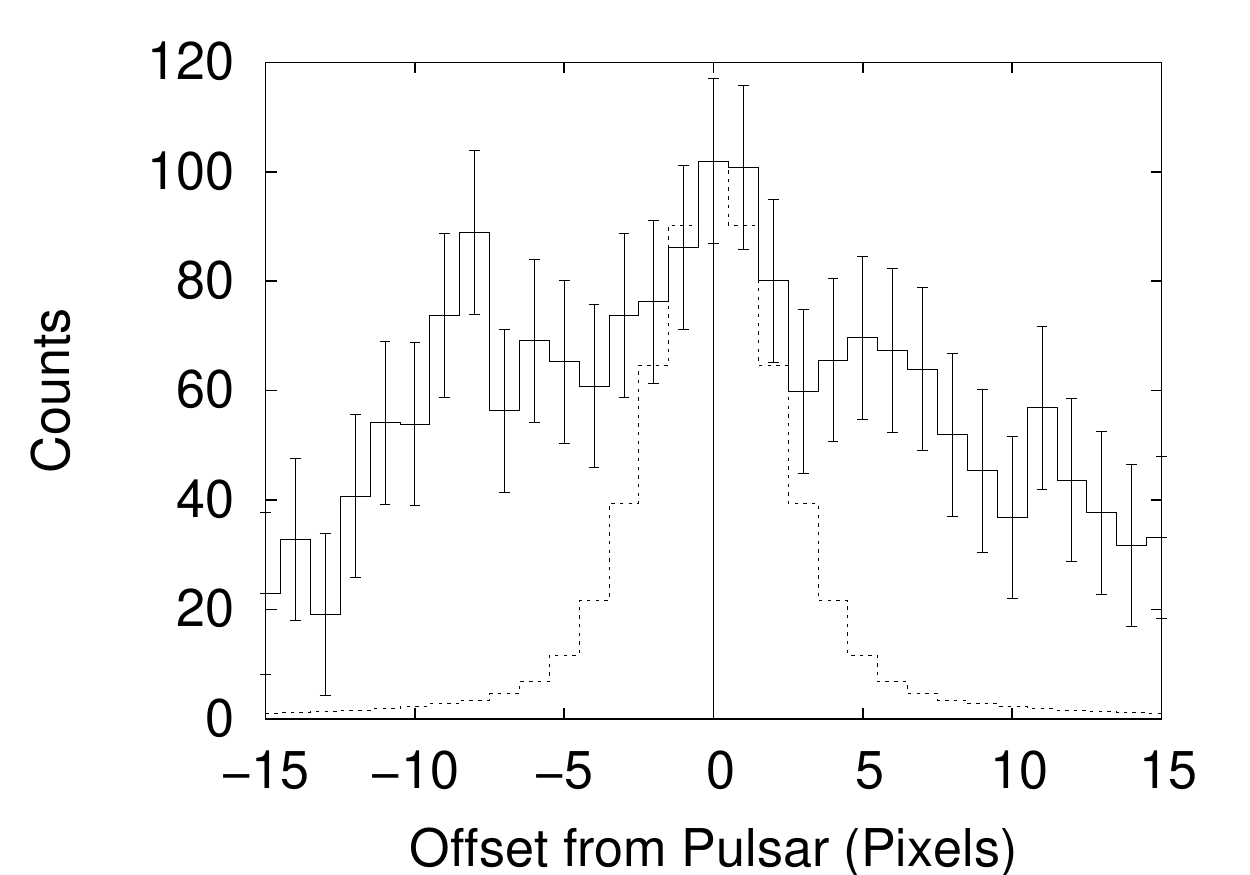}
\includegraphics[width=70mm]{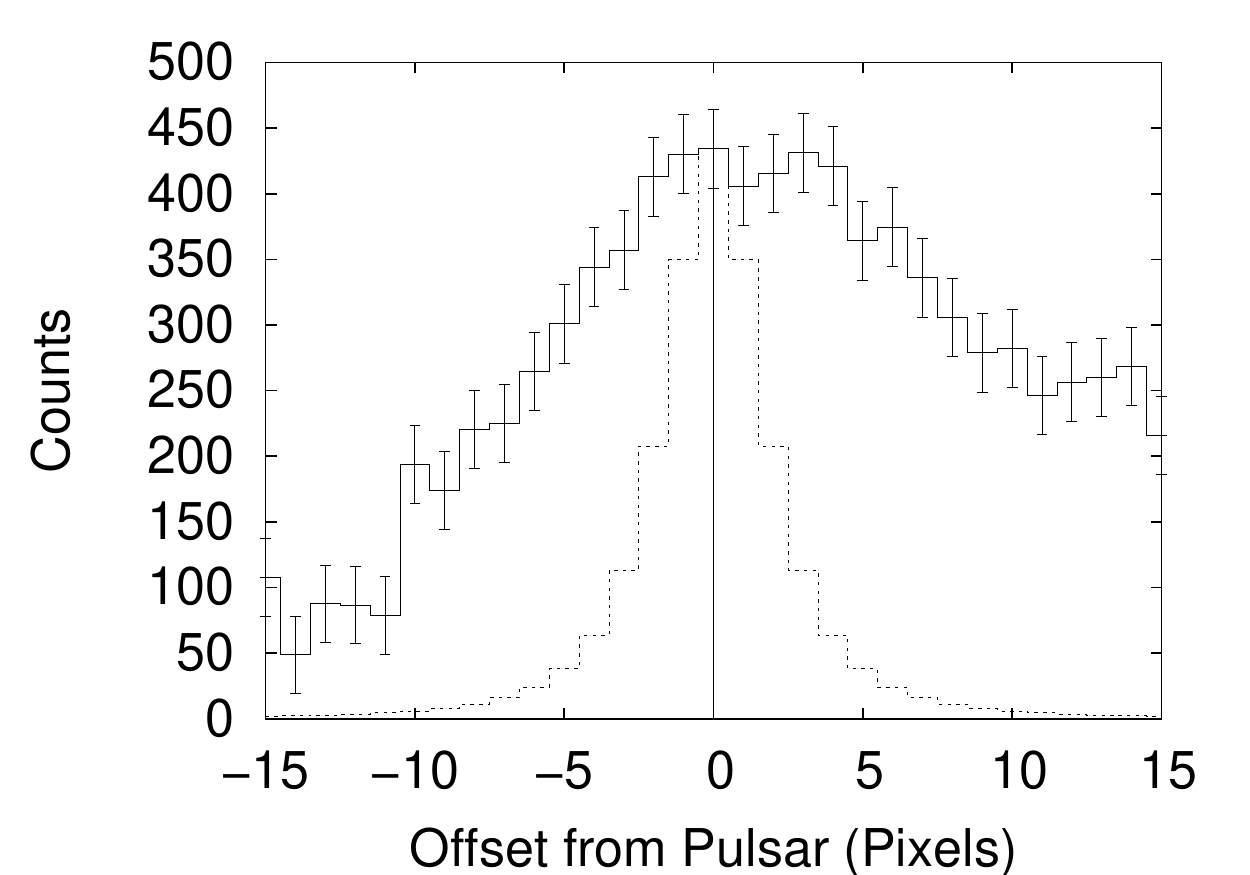}
\includegraphics[width=70mm]{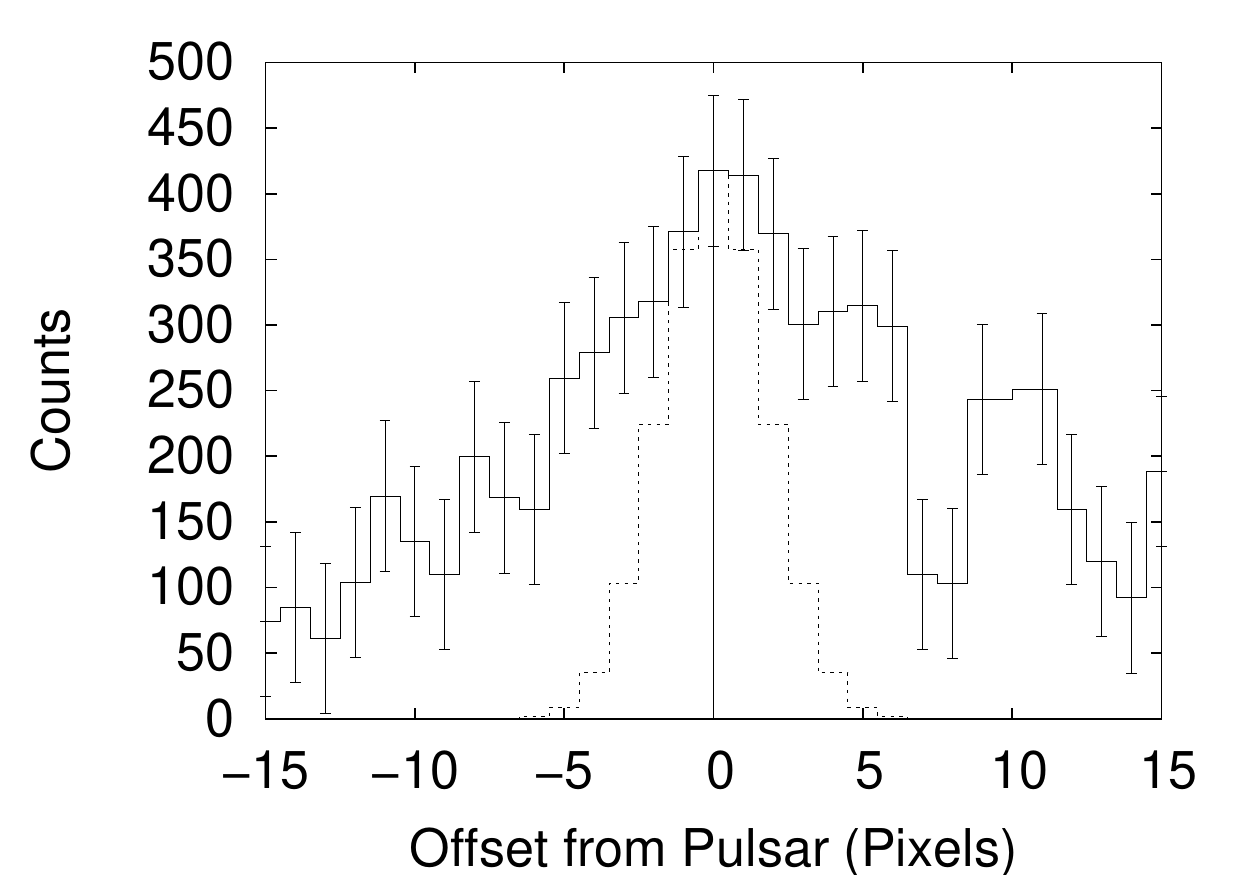}
\caption{ {\em Gemini} g' (top), r' (middle), and i'-band (bottom) spatial brightness profiles of the PWN (solid histogram)  obtained from averaging the counts in a slice of 5 pixels width centred on the pulsar's radio position and aligned along the NS direction. Pixel values increase NS. The brightness structure of the PWN is clearly that of an extended source, whereas the peaks of the g' and i'-band brightness profiles of the object detected at the \psr\ position (solid vertical line) match their respective image PSFs (dashed histogram). This is not the case for the r'-band brightness profile due to the object's much lower detection level (3.6$\sigma$)}
\label{sbp}
\end{center}
\end{figure}

Figure 5 shows the optical contours overlaid upon the Gemini images, with a point-like object recognised on top of the central brightness maximum. The object's brightness profile is  that expected for a star of comparable faintness embedded in a bright nebula. Indeed,  its point spread function (PSF) is consistent with that  of the Gemini image, after accounting for the low signal--to--noise of the detection, seeing variations during the observations, and the stacking of multiple, dithered exposures, which blurs the resulting intensity profile of faint sources. 
This is shown in Figure \ref{sbp}, where we plotted the object's brightness profile in the {\em Gemini} g', r', and i'-bands. These were obtained from averaging the counts in a slice of 5 pixels width centred on the pulsar's radio position and aligned along the NS direction, so as to avoid the contribution of the bright star detected East of the nebula. As seen, while the brightness structure of the PWN is clearly that of an extended source,  the g' and i'-band brightness profiles of the object detected at the \psr\ position match reasonably well their respective image PSFs. This is not the case for the r'-band brightness profile due to the object's much lower detection level (3.6$\sigma$). Deep, high-spatial resolution observations will better separate the object profile from the surrounding bright PWN and better determine its morphology.  In this respect, our case may be similar to that of PSR\, B0540$-$69, also embedded in a bright PWN (e.g., De Luca et al.\ 2007),  for which both the morphology and intensity profile of its optical counterpart  are also not very well defined in ground-based non-adaptive optics images (e.g., Caraveo et al.\ 1992).

Thus, accounting for all caveats, our image analysis suggests that the point-like object detected at the {\em Chandra} and VLBI  \citep{2013...1302.5625} positions  is a stellar object and, as such, is probably associated with the pulsar. Indeed, the chance coincidence probability that an unrelated point source falls within the radio
error circle of \psr\ is, after accounting for the accuracy of the optical astrometry,  only 
$\sim 3 \times 10^{-4}$,
computed based on the density of stellar objects in the Gemini field $\rho \sim 0.0025$ arcsec$^{-2}$ as $1-\exp(-\pi\rho r^2)$. We consider such a probability low enough to rule out a chance coincidence association.

We also investigated the possibility that this object is not the pulsar but an emission knot in the PWN. Optical emission knots can be observed in some of the PWNe as the result of a different density of relativistic particles in the nebula caused, for instance, by the formation of shocks at the termination front of the pulsar beam and/or equatorial wind, or turbulent motions in the nebula. An emission knot was seen, e.g. in the PWN around PSR\, B0540$-$69 (De Luca et al.\ 2007). However, the knot was much fainter than the pulsar and located far off from the geometrical centre of the PWN, as expected according to its possible formation mechanisms, whereas in the case of \psr\  it would be virtually coincident with the pulsar's position. An emission knot very close to the pulsar ($\sim 0\farcs6$) has been seen in the Crab PWN (see Moran et al.\ 2013 and references therein).  However, also in this case the knot is much fainter than the pulsar. Thus, we tend to conclude that it is unlikely that the object detected at the {\em Chandra} and VLBI positions is associated with a knot in the PWN, although we cannot completely rule out this possibility. Future observations of the PWN aimed at searching for possible flux variability from this object will help to confirm our conclusion.

Thus, on the basis of the positional coincidence of the object  detected in the Gemini images with the accurate {\em Chandra} and VLBI coordinates and the centre of symmetry of the optical PWN in the Gemini image, we conclude that it is a plausible optical counterpart to \psr.

\subsection{Photometry}

We performed aperture photometry of the \psr\ candidate counterpart to determine its flux in the Gemini g', r', and i'-band images and compare it with the upper limits derived in the INT, TNG, GTC, and {\em HST} images. We used the {\sc IRAF} routine {\sc daofind} to search for and measure the flux of potential point sources detected at the pulsar's position. In all cases we used an aperture of diameter comparable with the image PSF, to minimise contamination from the bright PWN.  In all cases, we applied the aperture correction computed from the growth curve of a number of unsaturated stars identified in the field. Details on the aperture photometry (aperture size, background annulus, etc.) and the applied aperture correction for each observation  are given in the corresponding sub-sections

\subsubsection{INT and TNG observations}

For the photometry calibration of the INT R and TNG V and R-band images  we used the set of five secondary photometric standards defined in \cite{2008MNRAS.390..235S} and identified directly in the frames. We matched the measured fluxes of these stars to their tabulated V and R-band magnitudes and produced a linear fit (with $\chi^2$=0.9999). We verified that the computed photometric zero-points are consistent with those tabulated in the instrument web pages\footnote{{\tt http://www.ing.iac.es/astronomy/instruments/wfc/}}\footnote{{\tt http://www.tng.iac.es/instruments/lrs/}}, which are 25.6, 26.2, and 26.3 for the INT R, TNG V and TNG R images, respectively.
We computed the 3-$\sigma$ limiting magnitudes of the images using the standard approach described in Newberry (1991). 
The 3-$\sigma$ limiting magnitudes for INT R, TNG V and TNG R images were 23.5, 24.7, and 24.3, respectively. For both the TNG and INT images we corrected for the airmass using the average atmospheric extinction terms for the La Palma Observatory from \cite{2003Obs...123..145K}.

\subsubsection{GTC observations}
 
Images of the Landolt standard star fields SA\, 95, SA\, 113, and G\,158 were taken  the same nights as the science observations and used for the photometry calibration. We verified the computed calibration against the photometric zero-points tabulated in the OSIRIS instrument web page\footnote{{\tt http://www.gtc.iac.es/instruments/osiris/}}, which are 29.2 and 28.8 for the r' and i'-band images, respectively. Using the same approach as described above, we placed 3$\sigma$ detection limits of 24.4 and 23.0  in the r' and i'-band images, respectively. Also in this case, we used the average atmospheric extinction terms from \cite{2003Obs...123..145K}.


\subsubsection{Gemini observations}

For the Gemini photometry calibration, we used the average airmasses, the tabulated zero-points and extinction coefficients reported in Jorgensen et al.\  (2009)\footnote{{\tt http://www.gemini.edu/sciops/instruments/gmos/}}. We then derived absolute zero-points of 27.9, 28.1, and 27.9 for the g', r', and i' images, respectively. We have cross-checked the photometry calibration against sets of secondary calibration stars identified on--the--frame and selected from our observations of the field performed with the GTC. 
Using {\tt daofind} we detected a source, at the  5.5$\sigma$, 3.6$\sigma$, and 5.5$\sigma$ level, in the Gemini g', r', and i'-band images, respectively, whose position is consistent with the error circle on the pulsar coordinates obtained from the {\em Chandra} and Green Bank VLBI observations. We then performed aperture photometry on this source. For each band, we used an aperture with a radius equal to 0\farcs5. We measured the sky background in an annulus of width 0\farcs3 located 0\farcs6 beyond the inner aperture. Both this annulus and the photometric aperture were chosen so as to reduce the contribution of the flux from the nebula. We then applied the computed aperture corrections to our photometry. This yielded magnitudes of $\rm g'=27.4\pm0.2$; $\rm r'=26.2\pm0.3$; and $\rm i'=25.5\pm0.2$ for the candidate pulsar counterpart.
 
\subsubsection{HST observations}
 
We applied the photometric calibration to the {\em HST} images by computing the count-rate  to flux  conversion  using  the  updated values of the keywords  {\tt PHOTFLAM}  and  {\tt PHOTPLAM} recorded in the image headers following the recipe described in the WFC3 Instrument and Data Handbooks (Dressel 2012; Rajan et al.\ 2010).  This yielded photometric zero-points of 25.5 and 24.8. We also applied the charge transfer efficiency (CTE) and aperture corrections according to the tabulated values in WFC3 Handbooks. The 3$\sigma$ detection limits above the sky background are $\sim$ 26.0 and $\sim$ 25.0 magnitude in the 625W and 775W images, respectively.

\begin{table*}
\caption{\psr\ coordinates: radio VLBI (Bietenholz et al. 2013), re-computed {\em Chandra} X-ray (this work), and candidate optical counterpart (Gemini data). Included are the error-circles. The confidence levels are all 90\%.}
\begin{center}
\begin{tabular}{|c|c|c|c|}
\hline
Observations		& RA (J2000)		& Dec. (J2000)						& Error-circle		\\ 
  			& h m s			&\degr \ \ $^{\prime}$ \ $^{\prime\prime}$		&$^{\prime\prime}$	\\ \hline
Radio VLBI		&02 05 37.920		&64 49 41.30					&0.2			\\
Chandra			&02 05 37.950		&64 49 41.60					&0.4			\\
Gemini	g'		&02 05 37.947		&64 49 41.17					&0.2			\\
Gemini	r'		&02 05 37.935		&64 49 41.31					&0.2			\\
Gemini	i'		&02 05 37.914		&64 49 41.42					&0.2			\\
\hline	
\end{tabular}
\end{center}
\label{coords}
\end{table*}

\begin{table*}
\caption{Summary of the photometry of the \psr\ candidate counterpart. All values are in the AB magnitude system (Oke \& Gunn 1983).  The errors quoted are purely statistical and do not include those due to the photometric calibrations. For the images where the pulsar is undetected, we quote the 3$\sigma$ detection limits on the sky background level (Newberry 1991). Fluxes have been corrected for the interstellar extinction using the coefficients of Fitzpatrick (1999) and the extinction value inferred from the hydrogen column density $N_{\rm H}$ best-fitting the {\em XMM-Newton} spectrum (Marelli 2012).}
\begin{center}
\begin{tabular}{|c|c|c|c|c|c|c|c|c|c|}
\hline
Telescope     & Filter     & Total Exposure & Average      &   Average	        & \multicolumn{2}{c|} {Observed Flux}        &Sig. ($\sigma$)         & \multicolumn{2}{c|} {Unabsorbed Flux} \\
                         &               & Time (s)           & Airmass      & Seeing               & mag.             & $\mu$Jy    &--                    & mag.             & $\mu$Jy\\
\hline
INT		&Harris R	&7320		&1.35        &1\farcs3				&23.5			&$<$ 1.45		&--			&$<$ 21.8		&$<$ 6.69\\
\\
TNG		&Johnson V	&6060		&1.5         &1\farcs3				&24.7			&$<$ 0.48		&--			&$<$ 22.5		&$<$ 3.63 \\
TNG		&Johnson R	&4830		&1.5         &1\farcs3				&24.3           	&$<$ 0.69     		&--			&$<$ 22.6		&$<$ 3.21 \\
\\
GTC		&r'		&700		&1.25        &0\farcs9				&24.4            	&$<$ 0.63     		&--			&$<$ 22.7		&$<$ 2.87\\
GTC		&i'		&2380		&1.25        &0\farcs9				&23.0            	&$<$ 2.29    		&--			&$<$ 21.8		&$<$ 6.96\\
\\
Gemini		&g\_G0301	&10920		&1.62        &0\farcs73			&$27.4\pm0.2$        	& $0.04\pm0.007$	&5.5		&$24.8\pm0.2$		&$0.44\pm0.08$\\
Gemini		&r\_G0303	&8220		&1.64        &0\farcs60			&$26.2\pm0.3$        	&$0.12\pm0.03$	&3.6		&$24.5\pm0.3$		&$0.58\pm0.16$\\
Gemini		&i\_G0302	&5521		&1.48        &0\farcs57			&$25.5\pm0.2$        	&$0.23\pm0.04$	&5.5		&$24.3\pm0.2$		&$0.69\pm0.13$\\
\\
HST		&625W        	&1800		&--		&--				&26.0            	&$<$ 0.14 		&--			&$<$ 24.3		&$<$ 0.69\\
HST		&775W        	&1950		&--		&--				&25.0            	&$<$ 0.36    		&--			&$<$ 23.8		&$<$ 1.10\\           
\hline
\end{tabular}
\end{center}
\label{results}
\end{table*}

\subsubsection{Summary}


Table \ref{coords} shows the coordinates of the candidate optical counterpart of \psr\ detected in the Gemini images. Also shown are the radio VLBI coordinates (Bietenholz et al. 2013) of \psr\ and the re-computed X-ray coordinates of the pulsar based upon the re-analysis of archival {\em Chandra} observations of 3C\,58. We also include the position error-circles, and confidence levels. The error-circles of the optical and X-ray positions, based upon the astrometry, are $\sim0\farcs2$ and $\sim0\farcs4$, respectively. The position of the candidate optical counterpart is consistent, within the error circles, with the radio VLBI and re-computed X-ray positions.

Table \ref{results} shows the results of our multi-band photometry for all the available image data set. The listed upper limits are at $3\sigma$. 
The reported spectral fluxes have been computed from standard formulae after converting the measured magnitudes to the AB system (Oke \& Gunn 1983).
We used the multi-band photometry to determine the slope of the pulsar's spectrum in the 5000-9000 \AA\ range.  To correct our flux measurements for the interstellar extinction towards \psr\ we used the value of the hydrogen column density derived from the X-ray spectral fits to the {\em Chandra} soft X-ray spectrum and the relation of \citet{1995A&A...293..889P}. 
 
Marelli (2012) analysed the full {\em Chandra} data set and best-fitted the soft X-ray spectrum with a two-component model consisting of a Power law (PL) ($\Gamma_{\rm X}=1.77\pm 0.03$) plus a blackbody (BB) with temperature $kT=1.88^{+0.14}_{-0.13}$ MK and an emitting radius $R_{\rm BB}=2.03^{+0.43}_{-0.36}$ km for a 3.2 kpc distance.  This yields a hydrogen column density, $N_{\rm H}= 4.5^{+0.13}_{-0.11} \times10^{21}$ cm$^{-2}$ and an $A_V = 2.5^{+0.07}_{-0.06}$. We note that \citet{Fesen08}, based upon spectroscopy studies of the 3C 58 SNR, derived an  E(B-V)=0.5--0.7 and  $A_V$ =1.6--2.3. Whereas  \citet{Fesen88} give E(B-V)=0.68$\pm$0.08. This corresponds to  $A_V=2.1^{+0.25}_{-0.25}$ for R=3.1.  These values are consistent with what we derived from the $N_{\rm H}$ at the $\sim 1.5 \sigma$ level. In the following, we assume the value of the extinction derived from the $N_{\rm H}$ obtained  from the best fit to the {\em Chandra} spectrum.  This also makes the comparison between the unabsorbed optical fluxes and the unabsorbed {\em Chandra} spectrum more consistent. Then, we applied the extinction correction in the different bands using the extinction coefficients of Fitzpatrick (1999). The extinction-corrected magnitudes and fluxes are reported in the last two columns of Table \ref{results}.

\section{Discussion}

\subsection{The Pulsar identification}

With the identification of \psr, the number of rotation-powered pulsars with either an identified or proposed optical counterpart (Mignani 2011) would amount to fourteen.  For ten of them, the identification is firmly secured either through the detection of optical pulsations at the radio period, or the tight positional coincidence with the radio position, or from the optical spectrum. The possible identifications of PSR\, J0108$-$1431 (Mignani et al.\ 2008) and PSR\, J1357$-$6429  (Mignani et al.\ 2011; Danilenko et al.\ 2012) are not yet confirmed, while that of PSR\, B1133+16 (Zharikov et al.\ 2008) seems now confirmed, albeit still marginally (Zharikov \& Mignani 2013).   \psr\  would be also one of  the very few  $\gamma$-ray pulsars identified in the optical. In particular,  it would be, possibly together with PSR\, J1357$-$6429, the first identified after the launch of {\em Fermi}, whereas most of the others have been identified after their detection by the {\em Compton Gamma Ray Observatory}.

One can speculate on whether our candidate counterpart can be identified or not with object o2 of Shearer \& Neustroev (2008), tentatively detected as an unresolved emission knot in the nebula in their 4.2m William Herschel Telescope (WHT) images, which they suggested as a possible pulsar counterpart. The same object was also proposed as the \psr\ counterpart by Bietenholz et al.\ (2013), based on the positional coincidence between its coordinates,  $\alpha =02^{\rm h}  05^{\rm m} 37\fs93$ and  $\delta  = +64^\circ 49\arcmin 41\farcs4$ (Shearer \& Neustroev 2008), and the updated {\em Chandra} and VLBI radio ones  (see Figure 5). However, we note that the flux of object o2 (R=24.15)  measured by Shearer \& Neustroev (2008) is much fainter than that of our candidate counterpart detected in the Gemini image (r'=$26.2\pm0.3$).  This means that, if object o2 were the pulsar's counterpart, its optical emission must have varied by about an order of magnitude.  According to the observed optical emission properties of rotation-powered pulsars, this is an unrealistic scenario. Alternatively, object o2 could have been a variable emission knot in the PWN, undetected in our R-band TNG images down to R$\sim 24.3$ (see Table \ref{results}) but detected at a much fainter flux level in the Gemini images.  However, the possibility that our counterpart is an emission knot in the PWN is unlikely (see our discussion in Sectn.  4.1), although we cannot completely rule it out.  Interestingly enough, objects o1 and o3 of Shearer \& Neustroev (2008) are also undetected in both the TNG and Gemini images (see Figure 5). This suggests that, if they also were emission knots in the PWN, they varied in flux of about the same amount as object o2 and, possibly, on the same time scale.  Thus, we are prone to believe that the detection of object o2 in the WHT images of Shearer \& Neustroev (2008), which was never confirmed in independent observations, was spurious.

\subsection{The Pulsar spectrum}

\begin{figure}
\begin{center}
\includegraphics[height=75mm]{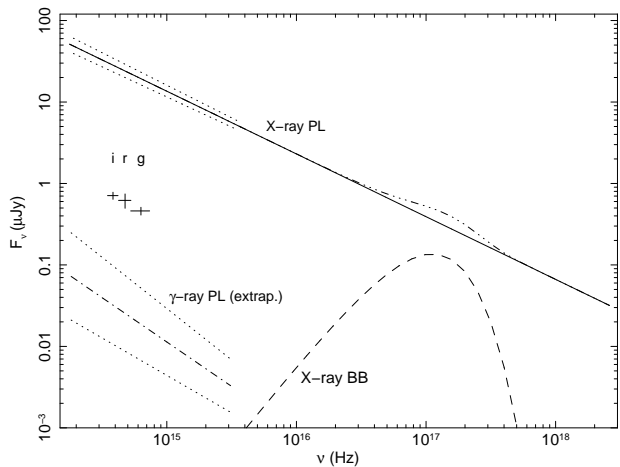}
\caption{Extinction-corrected optical (g',r',i') spectral flux measurements of the \psr\  candidate counterpart (data points). Optical flux upper limits in the other bands are not plotted, since they are above the g',r',i' measurements. Optical fluxes are compared with the unabsorbed pulsar X-ray spectrum measured by {\em Chandra} (Marelli 2012) and the extrapolation in the optical domain of its PL component (solid line), with photon index $\Gamma_{X} = 1.77\pm0.03$. The dotted lines correspond to the $1\sigma$ uncertainty on the PL extrapolation. For completeness, we also plotted the BB component (dashed line), with temperature $kT=1.88^{+0.14}_{-0.13}$ MK (emitting radius $R_{\rm BB}=2.03^{+0.43}_{-0.36}$ km at 3.2 kpc). The dot-dashed line is the extrapolation of the PL $\gamma$-ray spectrum of Abdo et al.\ (2010), with photon index $\Gamma_{\gamma}=2.09\pm0.17$ and exponential cut off at $E_{C} = 3.5\pm1.4$ GeV. Also in this case, the dotted lines correspond to the $1\sigma$ uncertainty on the PL extrapolation.}
\label{sed}
\end{center}
\end{figure}

We compared our extinction-corrected optical flux measurements of the \psr\ candidate counterpart with the low-energy extrapolations of the soft X-ray and $\gamma$-ray spectra measured by {\em Chandra} and {\em Fermi}, respectively.  In the hard X-rays (2.5--54 keV), the {\em RXTE} spectrum is described by a flatter PL with photon index $\Gamma_{\rm X}=1.06\pm 0.03$ (Kuiper et al.\ 2010) which, however, was obtained by fixing the hydrogen column density  $N_{\rm H}=3.4 \times10^{21}$ cm$^{-2}$, lower than derived from the fits to the {\em Chandra} spectrum (Marelli et al.\ 2011; Marelli 2012). We do not include the X-ray spectrum from Kuiper et al.\ (2010) since it has been computed for the pulsed component only and, as such, it is not directly comparable to the {\em Chandra} spectrum and the optical fluxes which are phase-averaged.  The $\gamma$-ray spectrum is described by a PL with photon index $\Gamma_{\gamma}=2.1\pm0.1\pm0.2$, where the first and second errors are statistical and systematic, and an exponential cut-off at an energy $E_{\rm C}=3.0^{+1.1}_{-0.7}$  GeV (Abdo et al.\ 2009). The $\gamma$-ray spectral parameters were slightly revised ($\Gamma_{\gamma}=2.09\pm0.17$; $E_{\rm C}=3.5\pm1.4$ GeV) in the First {\em Fermi} Catalogue of $\gamma$-ray pulsars (Abdo et al.\ 2010), although they are consistent with the previous ones.

Figure \ref{sed} shows the extinction-corrected optical fluxes of  the PSR\, J0205+6449 candidate counterpart compared with the multi-wavelength spectral energy distribution (SED). The Gemini fluxes lie well below the extrapolation of the  X-ray PL (Marelli 2012), hinting at the presence of a double break in the optical--to--X-ray spectrum, as observed in other rotation-powered pulsars (Mignani et al.\ 2010a).  In particular, the case of \psr\ is similar to that of PSR\, B0540$-$69, where a double break  is clearly present (Serafimovich et al. 2004; Mignani et al.\ 2010a; 2012a). Such a double break in the optical--to--X-ray spectrum would be at variance with what is observed for the Crab and Vela pulsars, where a single break is, instead, required to join the optical and X-ray fluxes. This would suggest a different energy and density distribution of relativistic particles in the neutron star magnetosphere, even for objects of comparable spin-down age.  The optical fluxes of  the \psr\ candidate counterpart can be described by a PL spectrum with photon index $\Gamma_{O} = 1.9 \pm 0.5$. This confirms that the optical emission from \psr\  would be non-thermal, as expected from its age (Mignani 2011). The value of the photon index of the optical PL would be comparable to those of most rotation-powered pulsars (Mignani et al.\ 2007; 2010a,b), for which the average values is $1.45$ with a $1\sigma$ scatter of 0.35, confirming that there is no obvious evolution as a function of the pulsar's age.  Once again, as in the case of PSR\, B0540$-$69 and also for \psr\ the value of the optical PL photon index would be compatible with that of the PL component of the {\em Chandra} X-ray spectrum ($\Gamma_{X}=1.77\pm 0.03$). Whether or not this suggests that the optical photons are related to the same population of relativistic electrons responsible for the non-thermal  X-ray emission, perhaps at different altitudes/latitudes in the neutron star's magnetosphere, is an interesting speculation. One way to address it may be by comparing the X-ray and optical light curves of \psr, which can be measured with the current generation of high-time resolution optical cameras, such as {\em IQueye} (Naletto et al.\ 2009).  The optical fluxes are well above the extrapolation of the $\gamma$-ray PL spectrum   (Abdo et al.\ 2010). This indicates the presence of an additional break in the pulsar spectrum, probably in the soft $\gamma$-rays/hard X-ray part. The discontinuity between the optical and $\gamma$-ray PL spectra is seen in other {\em Fermi} pulsars, such as PSR\, J0007+7303 (Mignani et al.\ 2013), while for others the available optical upper limits are too high with respect to the extrapolation of the $\gamma$-ray PL extrapolation to constrain the presence of a spectral break (Mignani et al.\ 2011; 2012b). It is only for PSR\,  J1048$-$5832 (Razzano et al.\ 2013) that the optical flux upper limits do not rule out that the optical spectrum is consistent with the extrapolation of the $\gamma$-ray PL. Thus, in general, the connection between the optical and $\gamma$-ray spectra in rotation-powered pulsars is unclear. Investigating such a connection is important to determine whether the emission in these two energy bands is related, as suggested by the fact  that both the pulsar optical and $\gamma$-ray luminosities seem to scale with the strength of the magnetic field at the light cylinder (Shearer \& Golden 2001).

\begin{table*}
\caption{Optical, X and $\gamma$-ray luminosity of all $\gamma$-ray pulsars with identified optical counterparts,  including \psr\ (in bold). The second and third columns list the spin-down age $\tau$ and the rotational energy loss $\dot{E}$, respectively, as listed in the ATNF pulsar data base (Manchester et al.\ 2005).  The distances and $\gamma$-ray luminosities ($>100$ MeV) are taken from Abdo et al.\ (2010). The X-ray luminosities (0.3--10 keV) are derived from the unabsorbed X-ray fluxes listed in Table 3.2 of Marelli (2012) and include the contribution of both thermal and non-thermal components.  The optical luminosities are computed from the pulsar magnitudes in the $V$ band, except for \psr\ (g') and PSR\, B1509$-$58 (R), and using the values of the associated interstellar extinction $A_{V}$ (Mignani 2011). The last three columns list the ratios between the optical, X-ray, and $\gamma$-ray luminosities, and the ratio between the optical luminosity and the rotational energy loss $\dot{E}$. }
\begin{center}
\begin{tabular}{|c|c|c|c|c|c|c|c|c|c}
\hline
Pulsar   &$Log(\tau)$ &$Log(\dot{E})$ & \multicolumn{3}{c|} {Log Luminosity  }  &d  & $Log(L_{opt}/L_X$) & $Log(L_{opt}/L_\gamma$)  & $Log(L_{opt}/\dot{E}$)  \\ 
              &    (yr)          &   (erg s$^{-1}$)   &                         &     		 (erg s$^{-1}$)   &                     &   (kpc)        &   \\  
              &                   &      & Optical  &  X-ray  &  $\gamma$-ray   &        		&                    &         &   \\ \hline 
Crab 				&3.10 & 38.65 &33.15		&36.32		&35.79 		&2		&-3.17		&-2.64	& -5.5 \\	
PSR\, B1509$-$58		&3.19 & 37.25 &30.97   	        &35.04      	&34.83     	         &4.2		&-4.07       	&-3.86	& -6.28 \\	
{\bf PSR\, J0205+6449}	&3.73 & 37.43 &30.06   		&33.42		&34.91     		&3.2		&-3.36      		&-4.85	& -7.37  \\   
Vela 			        		&4.05 & 36.84 &28.13		&32.44		&34.93		&0.287	&-4.31		&-6.80	& -8.71 \\	
PSR\, B0656+14		&5.05 & 34.58 &27.53		&32.22		&32.49		&0.288	&-4.69		&-4.96	& -7.05 \\	
Geminga				&5.53 & 34.51 &27.20     	        &30.97   		&34.39   		&0.250	&-3.77  		&-7.19	& -7.31 \\	
PSR\, B1055$-$52		&5.73 & 34.48 &28.20    	        &32.18    		&34.23     		&0.72	&-3.98       	&-6.03	& -6.28 \\	
\hline
\end{tabular}
\end{center}
\label{comp1}
\end{table*}

\subsection{The Pulsar luminosity}


We computed the ratio between the X-ray and optical fluxes of \psr\ from the available {\em Chandra} and Gemini measurements.  The spectral fit of Marelli  (2012), based on a PL plus BB model, gives  an unabsorbed  X-ray flux $F_{\rm X} = (19.7\pm 0.7) \times 10^{-13}$ erg cm$^{-2}$ s$^{-1}$ in the 0.3--10 keV energy range for the PL component only, whereas the total X-ray flux is $F_X = (21.8\pm0.8) \times 10^{-13}$ erg cm$^{-2}$ s$^{-1}$\footnote{We note that the value reported in Table 3.2 of Marelli (2012) is affected by a typo (Marelli, private communication).}. Hereafter, we assume the total X-ray flux as a reference. On the basis of the Gemini g' detection, and using the same extinction correction as above, we computed the unabsorbed flux of the pulsar in the g' band would be $F_{g'}=  9.43\times 10^{-16}$ erg cm$^{-2}$ s$^{-1}$.  By assuming a distance of 3.2 kpc, the unabsorbed fluxes give g'-band and X-ray luminosities of $L_{g'} = 1.15 \times 10^{30}$ erg s$^{-1}$ and $L_{X} = 2.66 \times 10^{33}$ erg s$^{-1}$. This gives an optical--to--X-ray luminosity ratio $L_{g'}/L_{\rm X}=4.32 \times 10^{-4}$. Similarly, we computed the unabsorbed optical--to--$\gamma$-ray luminosity ratio. As a  reference, we assumed  the $\gamma$-ray flux above 100 MeV measured by {\em Fermi}, $F_{\gamma}= (6.64 \pm 0.65) \times 10^{-11}$ erg cm$^{-2}$ s$^{-1}$ (Abdo et al.\ 2010).  For the assumed distance of 3.2 kpc the $\gamma$-ray luminosity of \psr\  is $L_{\gamma} = 8.13 \times 10^{34}$ erg s$^{-1}$. This gives a luminosity ratio $L_{g'}/L_{\gamma} =1.41 \times 10^{-5}$.  The computed optical, X-ray, and $\gamma$-ray luminosities of \psr, together with their ratios, are listed in Table\, 3, where they are compared with the corresponding values measured for the other $\gamma$-ray pulsars identified in the optical. As a reference, for all pulsars we assumed the same pulsar distances as used  in Abdo et al.\ (2010), the unabsorbed X-ray fluxes in the 0.3--10 keV energy range listed in Marelli (2012), and the $\gamma$-ray fluxes above 100 MeV listed in  Abdo et al.\ (2010).  We computed the optical luminosities according to the observed V-band magnitudes and extinction $A_V$ (Mignani 2011).  As seen, \psr\  would be the $\gamma$-ray pulsar with the second highest optical luminosity after the Crab pulsar, as expected from its low spin-down age ($\tau \sim$ 5400 year). The optical--to--X-ray and optical--to--$\gamma$-ray luminosity ratios of \psr\ would be consistent with those measured for other young pulsars ($\tau \la 10$ kyr), with the Crab pulsar being that with the highest ratios. However, the luminosity ratios for \psr\ would be larger than those of the slightly older Vela pulsar ($\tau \sim 11$ kyr), owing to the fact that the latter is about two orders of magnitude fainter in the optical. 

We also computed the ratio between the derived optical luminosity $L_{g'} = 1.15 \times 10^{-30}$ erg s$^{-1}$ of \psr\ and its rotational energy loss $\dot{E}=2.7 \times 10^{37}$ erg s$^{-1}$. Assuming a 3.2 kpc distance, we obtained  $L_{g'}/\dot{E}=4.27 \times 10^{-8}$.  Again, in Table \ref{comp1} this value is compared with those of the other $\gamma$-ray pulsars identified in the optical domain. Interestingly enough, the optical efficiency for \psr, defined as $\eta_{opt} \equiv L_{opt}/\dot{E}$, would be lower than for the other young pulsars (Crab and PSR\, B1509$-$58) but higher than the Vela pulsar, confirming a trend for a decrease in the optical emission efficiency of young pulsars as a function of the spin-down age.  The existence of such a trend has been already proposed (e.g. Zharikov et al.\ 2006), but, so far, the assumption only relied on the computed lower emission efficiency of the Vela pulsar with respect to the other young pulsars Crab, PSR\, B1509$-$58, and PSR\, B0540$-$69, with the latter ($\eta_{opt} =1.07 \times 10^{-5}$) not listed in Table 3 because it has not been yet detected in $\gamma$-rays. Thus, it has been unclear whether the optical emission efficiency of Vela-like pulsars was indeed intrinsically lower than the Crab-like ones, or Vela stood out as a peculiar case. The probable optical identification of the $\sim$5400 year-old \psr, which is ideally half way between the two classes,  now represents an important piece of evidence in favour of this interpretation. The upper limits on the optical luminosities of Vela-like $\gamma$-ray pulsars, such as PSR\, B1706$-$44, PSR\, J1357$-$6429, PSR\, J1028$-$5819, 	PSR\, J0007+7303 (Mignani et al.\ 1999; 2011; 2012b; 2013) and  PSR\, J1048$-$5832 (Razzano et al.\ 2013), are consistent with this trend. Of course, the optical identification of some of these pulsars will eventually give the long sought proof.  Such a trend can be interpreted as the result of the secular decrease of the pulsar non-thermal optical luminosity, an effect predicted by Pacini \& Salvati (1983) as a result of the pulsar spin down.  We note that the optical emission (non-thermal) efficiency tends to increase again for the middle-aged pulsars (PSR\, B0656+14 and Geminga).

\section{Summary and Conclusions}

We performed multi-band optical observations of \psr\ with a variety of ground-based facilities, including the 8m Gemini and 10.4m GTC telescopes, and the {\em HST}. We detected a  possible candidate counterpart to the pulsar, with an i'-band magnitude of  25.5 (5.5$\sigma$ detection), based upon its positional coincidence with the recently measured {\em Chandra} and radio coordinates (Bietenholz et al.\ 2013). Thus, \psr\ would possibly be the fourteenth pulsar with either an identified or proposed optical counterpart (Mignani 2011), and the eighth of the $\gamma$-ray pulsars detected by {\em Fermi}. The pulsar's spectrum  would be consistent with a PL with photon index $\Gamma_{O} = 1.9 \pm 0.5$, similar to the X-ray one ($\Gamma_{X}=1.77\pm 0.03$).  The multi-wavelength SED of \psr\ show that the optical fluxes would lie below and above the extrapolations in the optical domain of the X and $\gamma$-ray  PL, respectively, indicating a break in the optical/X-ray region, after that in the X/$\gamma$-ray. The optical luminosity,  $L_{opt} = 1.15 \times 10^{-30}$ erg s$^{-1}$, would imply an emission efficiency $\eta_{opt}=4.27 \times 10^{-8}$, higher than the two times older Vela pulsar, confirming a trend for a decrease of the optical emission efficiency with the spin-down age in young pulsars. Due to the presence of a bright PWN, deeper high spatial resolution observations with the {\em HST} are better suited to resolve the presence of a point source, confirm the optical identification of \psr, and better study its spectrum. 
Moreover, monitoring for  changes in the flux of the candidate counterpart on a few month/year time scale would allow one to rule out the possibility that it is an emission knot in the PWN.

%

\section*{Acknowledgments}
Based on observations made with the Gran Telescopio Canarias
(GTC), installed in the Spanish Observatorio del Roque de los
Muchachos of the Instituto de Astrofsica de Canarias, in the island
of La Palma. The research leading to these results has received
funding from the European Commission Seventh Framework
Programme (FP7/2007-2013) under grant agreement n.
267251. PM is grateful for his funding from the Irish Research Council (IRC). We are grateful to the anonymous referee whose comments improved the quality of the final version of the paper.

\label{lastpage}

\end{document}